\documentclass[journal]{IEEEtran}
\usepackage{amsmath,amssymb,amsfonts}
\usepackage{subfigure}
\usepackage{graphicx}
\usepackage{caption}
\usepackage{diagbox}
\usepackage{color}
\usepackage{makecell}
\usepackage{cite}
\usepackage{stfloats}

\begin{document}

\title{Reconfigurable Intelligent Surface: \\ Power Consumption Modeling and Practical Measurement Validation}

\author{Jinghe~Wang, Wankai~Tang, Jing Cheng Liang, Lei~Zhang, Jun Yan Dai,\\ Xiao~Li, Shi~Jin, Qiang~Cheng, and~Tie Jun~Cui
	
\thanks{Jinghe Wang, Wankai Tang,  Xiao Li and Shi Jin are with the National Mobile Communications Research Laboratory, Southeast University, Nanjing 210096, China  (e-mail: wangjh@seu.edu.cn; tangwk@seu.edu.cn; li\_xiao@seu.edu.cn; jinshi@seu.edu.cn).

Jing Cheng Liang, Lei Zhang, Jun Yan Dai, Qiang Cheng, and Tie Jun Cui are with the State Key Laboratory of Millimeter Waves, Southeast University, Nanjing 210096, China  (e-mail: jingcheng\_liang@foxmail.com; cheunglee@126.com; junyand@seu.edu.cn; qiangcheng@seu.edu.cn; tjcui@seu.edu.cn).

}}

\maketitle

\begin{abstract}
The reconfigurable intelligent surface (RIS) has received a lot of interest because of its capacity to reconfigure the wireless communication environment in a cost- and energy-efficient way.
However, the realistic power consumption modeling and measurement validation of RIS has received far too little attention. Therefore, in this work, we model the power consumption of RIS and conduct measurement validations using various RISs to fill this vacancy. Firstly, we propose a practical power consumption model of RIS. The RIS hardware is divided into three basic parts: the FPGA control board, the drive circuits, and the RIS unit cells. The power consumption of the first two parts is modeled as  $P_{\text {static}}$ and that of the last part is modeled as $P_{\text {units}}$. Expressions of $P_{\text {static}}$ and $P_{\text {units}}$ vary amongst different types of RISs. 
Secondly, we conduct measurements on various RISs to validate the proposed model. Five different RISs including the PIN diode, varactor diode, and RF switch types are measured, and measurement results validate the generality and applicability of the proposed power consumption model of RIS. Finally, we summarize the measurement results and discuss the approaches to achieve the low-power-consumption design of RIS-assisted wireless communication systems.

\end{abstract}


\IEEEpeerreviewmaketitle

\section{Introduction}

\IEEEPARstart{W}{ireless} communication has been evolving rapidly over the past few decades, and it is expected to continue to advance and transform the way we communicate, interact, and work in our society. Some of the key trends and developments including extremely-high data rates, improved reliability and coverage, enhanced security and privacy, etc., are driving the sixth generation (6G) dedicated to introducing a variety of developing technologies to shape brand-new wireless communication paradigms. There are several promising technologies that are being explored and developed, such as millimeter-wave (mmWave) and terahertz (THz) communication, artificial intelligence (AI) and machine learning (ML), heterogeneous networks (HetNets), and so on. As the demand for wireless networks grows rapidly, however, those promising technologies must deal with increasingly complicated networks, expensive hardware, or rising energy usage. Taking the mmWave communication as an example, the mmWave band suffers severe atmosphere propagation attenuation, thus having a short transmission distance. Therefore, the corresponding beamforming technologies are needed to generate strong directional narrow beams for the user equipment (UE) to overcome serious path loss. The full digital beamforming, nevertheless, is quite expensive and high-energy consuming because of using a large number of RF chains. The hybrid architecture can significantly reduce the hardware complexity and energy consumption of the system, however, it has a significant computational complexity due to the complicated hybrid beam precoding methods. Accordingly, the practical application of mmWave communication is still challenging.

Recently, the reconfigurable intelligent surface (RIS) becomes much more popular in wireless communication due to its ability to intelligently control large-scale low-cost passive unit cells in a low-power manner to reshape the wireless environment, which has gained the high attention of the wireless research community. Specifically, RIS is a two-dimensional reconfigurable planar which developed from metamaterials \cite{ma2020information,cui2017information}. It consists of a large number of small and low-cost electromagnetic elements, called “RIS unit cells”. These RIS unit cells are typically made of metals, media, and adjustable components (i.e., PIN diodes, varactor diodes, RF switches, etc.). In particular, adjustable components help modify the behavior of electromagnetic waves by adjusting the electromagnetic parameters such as the phase, amplitude, and polarization \cite{cui2014coding,zhang2019breaking}, thus enabling the wireless propagation environment to change from passive adaptation to active control \cite{9140329}. Moreover, RISs can be deployed in various scenarios, such as indoor environments, urban areas, and satellite communication systems.
Recently, academic research on RIS has expanded from sub-6 GHz to millimeter wave (30GHz-300GHz) and even to terahertz, since terahertz meta-elements can be useful in terahertz applications such as wireless communication, sensing, and imaging. The authors in \cite{natureCMOS} proposed a modular approach to create programmable terahertz metasurfaces using fully integrated silicon chip tiles. The devices can act as efficient switches even at terahertz frequencies by utilizing local resonances within the meta-element.

One of the main benefits of RIS-assisted wireless networks is to control the propagation of electromagnetic waves, and reflect and focus them toward specific directions, which helps reduce the impacts of obstructions and strengthen the received signal strength at the receiver. Especially when there is no direct link between the base station (BS) and the UE, RIS can be deployed to provide additional virtual line-of-sight (vLoS) links for maintaining the signal transmission between them. Also, system performance enhancement can be achieved through the joint active beamforming at the BS and passive beamforming (phase shift design) at the RIS. 
Theoretical studies of joint beamforming optimization were presented by many authors over the past years \cite{han2019large,wu2019intelligent,huang2019reconfigurable,guo2020weighted,8247211}. These works either significantly enhance the system's spectral/energy efficiency, sum rate, and minimum throughput, or minimize the total transmit power at the BS. 
Particularly in \cite{9989438}, the authors reported the latest research on the use of RIS and UAVs to maximize data collecting for future IoT networks. The UAV-mounted RIS can help with signal beamforming via its reflecting elements, boosting network dependability without increasing sensor transmission power and keeping the UAV's power consumption low.
Moreover, the path loss modeling, RIS prototyping system design and measurement validations were illustrated in \cite{tang2020wireless,9551980,9837936,10100676,dai2020reconfigurable,trichopoulos2022design,9765815}. These measurement results contribute to paving the way for practical RIS deployments and applications in the future. In addition, RIS can be reconfigured dynamically by artificial intelligence (AI) to adapt to changes in the wireless environment \cite{feng2020deep,yang2020deep,sheen2021deep,wang2021interplay,huang2020reconfigurable}.

Another advantage of RIS-assisted wireless networks is that they use less power than traditional amplify-and-forward (AF) and decode-and-forward (DF) relaying systems. Relays are often thought of as active devices that require a specific power source to operate and are used to transmitting the signals \cite{9119122}. Power amplifiers (PAs) and low-noise amplifiers (LNAs) are expensive and power-consuming electrical components required for transmission and reception, respectively. Dedicated PAs, in contrast, are usually not required for operating RISs. RISs can be essentially passive, with only a small amount of energy required to configure and enable the RIS unit cells. Concerning the power-consuming budget, it is commonly assumed in relay-assisted systems that the total system power consumption is allotted between the transmitter and the relay to ensure a total power constraint. Nonetheless, because the hardware structure of RIS and relay is fundamentally different, the power consumption of RIS and the power allocation of the RIS-assisted system must be thoroughly reconsidered.

The following works have studied or mentioned the power consumption of RIS \cite{huang2019reconfigurable,tang2020wireless,9551980,zappone2020overhead,9734027,dai2020reconfigurable}. 
In \cite{huang2019reconfigurable}, the authors developed a total power consumption model of the RIS-assisted system, which helps formulate and solute the energy efficiency maximization problem for the joint active and passive beamforming design. Based on the power consumption of the conventional phase shifter in RF circuits \cite{ribeiro2018energy,mendez2016hybrid,huang2018iterative}, the authors in \cite{huang2019reconfigurable} claim that the RIS power consumption depends on the number of RIS unit cells and the resolution of its individual unit cell. And the power consumption of RIS is modeled as $P_{\text{RIS}}=N P_n(b)$, where $N$ is the number of identical RIS unit cells and $P_n(b)$ denotes the power consumption of each RIS unit cell having $b$-bit resolution. However, both the power consumption of the control circuit and the effect of unit cell coding on power consumption are not considered, and the specific relationship between $b$ and power consumption is not clearly given. In \cite{zappone2020overhead}, the authors develop a total power consumption model, in which the process and the power consumption of the channel state information (CSI) estimation and feedback are additionally considered compared with \cite{huang2019reconfigurable}. Nevertheless, this work still mainly focuses on solving the optimization problems for the phase shift design of RIS, system transmission and reception, channel estimation, and feedback, which did not illustrate the power consumption model of RIS in depth. Besides, the authors compared the system performance of active RIS and passive RIS under the same overall power budget in \cite{9734027}. The overall power consumption of passive RIS-aided systems is given. The power consumption of RIS is modeled as $P_{\text{RIS}}=N P_{\text{SW}}$, where $N$ is the number of identical RIS unit cells and $P_{\text{SW}}$ denotes the power consumption of each RIS unit cell. Nevertheless, the $P_{\text{RIS}}$ in \cite{9734027} is still not very comprehensive since the power consumption of RIS's controller is not considered and the power consumption of each unit cell is still regarded as a constant. Specific measurements of RIS power consumption are mentioned in \cite{tang2020wireless,9551980,dai2020reconfigurable}. The authors in \cite{tang2020wireless} illustrate some preliminary measurement results of RIS power consumption, which show that different kinds of RIS have different power consumption characteristics. However, it has not been further modeled and discussed. In \cite{9551980}, a detailed description of the power consumption of the fabricated varactor-diode-based RIS is shown, which claims that this kind of RIS consumes little power, showing the great potential of RIS for future communication systems. In \cite{dai2020reconfigurable}, the authors gave the specific power consumption of the RIS as about 153 W when the total radiated power is 64 W, and revealed that the proposed RIS in \cite{dai2020reconfigurable} can reduce the power consumption effectively. Nevertheless, the specific power consumption model of RIS is undefined.

So far, there has been limited study on RIS power consumption modeling and practical measurement validation, inspiring our work to fill this void. In this work, a general power consumption model of RIS is proposed, and practical power consumption measurements of various RISs are conducted for validating the model. The establishment of a practical RIS power consumption model helps accurately characterize the power consumption of the system and energy efficiency performance of the RIS-assisted wireless communication system and provides a significant basis for subsequent wireless system optimization design and algorithm performance evaluation. In the following work, we focus on the power consumption research of RIS. We summarize the main contributions of this work as follows:

\begin{itemize}
\item Based on the observation and analysis of various fabricated RISs, we split the RIS hardware into three basic parts: the FPGA control board, the drive circuits, and the RIS unit cells. 
The total power dissipated to operate the RIS is composed of two parts, one is the power consumption of the first two parts, modeled as $P_{\text {static}}$ and the other is that of the last part, modeled as $P_{\text {units}}$.
\item The static power consumption model of RIS $P_{\text {static}}$ is first proposed: the power consumption of the FPGA control board is regarded as a constant value, however, that of the drive circuits is affected by the number of control signals and the inherent power consumption characteristics of individual drive circuits. In conclusion, the expression of the static power consumption $P_{\text {static}}$ is given.
\item The power consumption model of RIS unit cells $P_{\text {units}}$ is proposed: For the PIN diode-based RIS, $P_{\text {units}}$ is affected by multiple factors, such as the polarization mode, the number of RIS unit cells, controllable bit resolution, working states, etc. For the varactor diode-based RIS, $P_{\text {units}}$ is almost negligible. As for the RF switch-based RIS, $P_{\text {units}}$ is affected by the polarization mode, the number of RIS unit cells, and the inherent power consumption of individual RF switch devices. In conclusion, the expression of the power consumption of RIS unit cells $P_{\text {units}}$ is given.
\item Expressions of $P_{\text {static}}$ and $P_{\text {units}}$ vary amongst different types of RISs. The overall power consumption model of RIS is validated by measuring the practical power consumption of various RISs, i.e., PIN diode-/varactor diode-/RF switch-based RIS. And measurement results validate the generality and applicability of the proposed power consumption model of RIS. 
\item Finally, we summarize the measurement results and discuss the approaches to achieve the low-power-consumption design of RIS-assisted wireless communication systems.
\end{itemize}

The rest of this article is organized as follows. Section II illustrates the power consumption model of the RISs. Section III shows the practical power consumption measurement and validation of RISs. Five fabricated RISs including PIN diode, varactor diode, and RF switch types are measured to provide practical measurement results. Section IV presents the discussion about the measurement results of fabricated RISs and the low-power-consumption design of the RIS-assisted systems. Section V brings this effort to a close.

\section{Power Consumption Modeling of RIS}
In this part, the power consumption of the RIS is modeled. We start with an introduction of RIS hardware and split the RIS hardware into three parts, and then model the power consumption of each part respectively.

\subsection{The Introduction of RIS hardware}

\begin{figure}
\centering
\includegraphics[height=3.5cm,width=8.5cm]{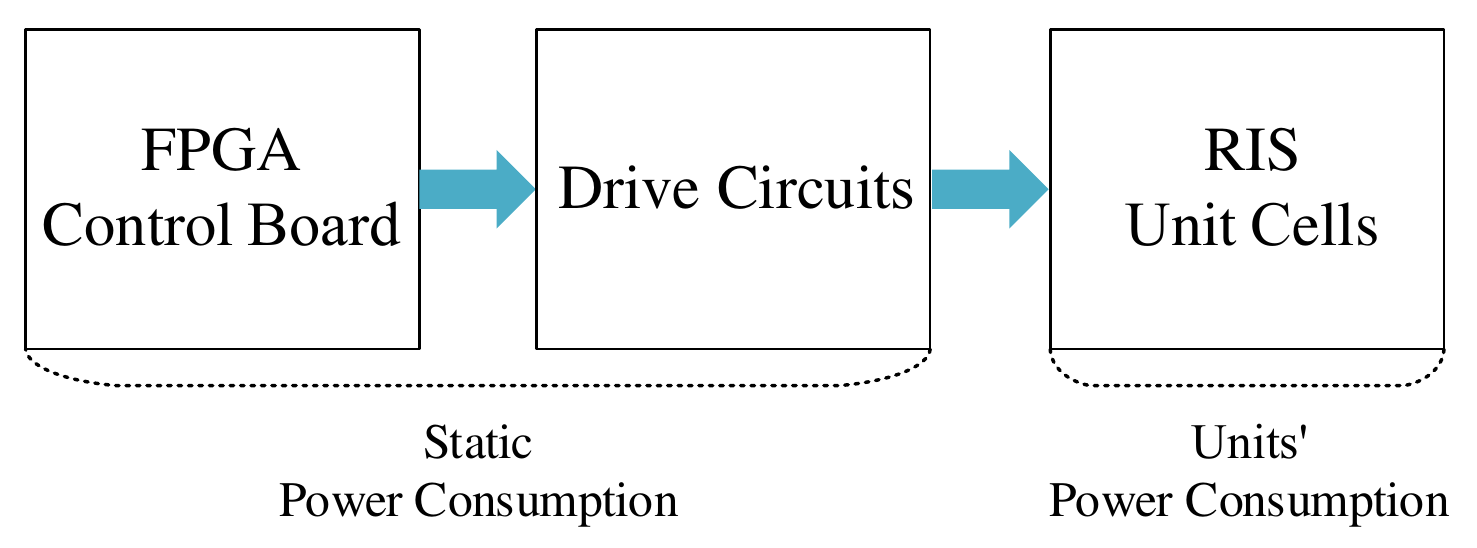}
\caption{General RIS hardware can be simply divided into three parts: the FPGA control board, drive circuits, and RIS unit cells.}
\label{1}
\vspace{-0.2cm}
\end{figure}

A RIS needs a controller. The controller (including an FPGA and drive circuits) is required for receiving the external signals, processing data and programming, as well as configuring the RIS unit cells. RIS unit cells usually consist of adjustable electronic components, metals, media, etc. Adjustable electronic components are one of the most significant elements whose electrical parameters (i.e., capacitance, inductance) can be changed in response to control signals. Tunable devices (i.e., PIN diodes, varactor diodes, and RF switches) can be used to construct programmable reflections to control electromagnetic waves. 
Overall, we split the RIS hardware into three basic parts. As shown in Fig.\ref{1}, general RIS hardware includes three parts: the FPGA control board, the drive circuits to drive adjustable electronic components, and RIS unit cells. 

%===========================================

%===========================================

Firstly, the FPGA control board is embedded to control and program the coding states of RIS unit cells. It is a programmable array of logic gates, which is not only used for providing clock signals but also for data processing, i.e., generating the corresponding RIS beamforming coding sequence configuration according to the preset algorithm based on the feedback data information. 
Secondly, the drive circuits are used to drive the adjustable electronic components. They are usually integrated circuits for generating variable current/voltage according to the control signals from the master FPGA  control board. Then, these adjustable electronic components are driven to achieve different coding states. 
Thirdly, the RIS unit cells are artificial electromagnetic elements that are composed of substrates, metallic vias, metal patches, and adjustable electronic components. They are regularly arranged on a two-dimensional surface and can generate different three-dimensional beamforming patterns according to the combination of different coding sequences.  
Overall, the total power consumption dissipated to operate the RIS consists of two parts, one is the static power consumption generated by the FPGA control board and drive circuits, and the other is the power consumption consumed by the RIS unit cells. Therefore, the total power consumption modeling of RIS can be expressed as follows:
\vspace{-0.1cm}
\begin{equation}
\large 
P_{\text{RIS}}=P_{\text {static }}+P_{\text {units}}
\end{equation}
where $P_{\text {static }}$ is the static power consumption and $P_{\text {units}}$ is the power consumption of RIS unit cells.

\subsection{The Static Power Consumption Modeling}

According to whether the RIS hardware power consumption is affected by the number of control signals, the static power consumption dissipated to operate the RIS is the superposition of two parts, one is the power consumption of the FPGA control board $P_{\text {control board}}$, and the other is the power consumption of drive circuits $P_{\text {total drive circuits }}$. The static power consumption modeling of RIS can be expressed as follows:
\begin{equation}
\large 
P_{\text{static}}=P_{\text {control board}}+P_{\text {total drive circuits }}
\end{equation}
Generally, an FPGA selected as the master control board needs to have enough arithmetic power to cover all data processing, thus $P_{\text {control board}}$ can be regarded as a constant value. However, $P_{\text {total drive circuits }}$ is variable that is related to the types of adjustable electronic components, the number of control signals, and self-power consumption characteristics.

\begin{itemize}
\item $P_{\text {total drive circuits }}$ is related to the types of adjustable electronic components since different components utilize different types of drive circuits. For example, PIN-diode-based RIS unit cells can be driven by shift registers; varactor-diode-based RIS unit cells can be driven by digital-to-analog converters (DACs) and operational amplifiers (op-amps), PWM signals + level regulator \cite{9551980}, or CMOS logic circuits; RF switch-based RIS unit cells can be driven by FPGAs and shift registers.
\item $P_{\text {total drive circuits }}$ is also related to the polarization mode $x=v,h$, in which $v$ represents the vertical polarization and $h$ represents the horizontal polarization. For dual-polarized RISs, drive circuits are required in each polarization direction to control the adjustable electronic components.
\item $P_{\text {total drive circuits }}$ is related to the number of control signals. Specifically, it is related to the number of adjustable electronic components $N_{\text {c},x}$ ($x=v,h$) and the control Degree-Of-Freedom (DoF) of RIS (i.e., unit cell control, row control, column control, and sub-array control). RIS unit cells divided into the same group can use the same control signals, and $N_{\text {g},x}$ ($x=v,h$) is denoted as the number of RIS unit cells that are in the same group with the same control signal.
\item Total drive circuits are composed of multiple individual drive circuits. Therefore, $P_{\text {total drive circuits }}$ is related to the self-power consumption characteristics of the individual drive circuit. $N_{\text {s}}$ is denoted as the number of control signals generated by each drive circuit and $P_{\text {drive circuit }}$ is denoted as the rated power consumption of a single drive circuit.
\end{itemize}

\begin{figure*}[ht]
\large 
\begin{equation}
P_{\text {static }}=\left\{\begin{array}{l}
P_{\text {control board }}+\left(\left\lceil\frac{\sum_{i=1}^N B_{i, v}}{N_{\text{g}, v} \cdot N_{\text{s}}}\right\rceil+\left\lceil\frac{\sum_{i=1}^N B_{i, h}}{N_{\text{g}, h} \cdot N_{\text{s}}}\right\rceil\right) \cdot P_{\text { drive circuit }}, \text { PIN-diode/RF switch-based RIS } \vspace{0.3cm}\\
P_{\text {control board }}+\left(\left\lceil\frac{\sum_{i=1}^N I_{i, v}}{N_{\text{g}, v} \cdot N_{\text{s}}}\right\rceil+\left\lceil\frac{\sum_{i=1}^N I_{i, h}}{N_{\text{g}, h} \cdot N_{\text{s}}}\right\rceil\right) \cdot P_{\text { drive circuit }}, \text { varactor-diode-based RIS }
\end{array}\right.
\end{equation}
%\hrulefill
\end{figure*}

\begin{figure*}[ht]
\large 
\begin{equation}
P_{\text {static }}=\left\{\begin{array}{l}
P_{\text {control board }}+\left(\mathbb{I}_{v}+\mathbb{I}_{h}\right)\left\lceil\frac{B \cdot N}{N_{\text{g}} \cdot N_{\text{s}}}\right\rceil \cdot P_{\text {drive circuit }}, \text { PIN-diode/RF switch-based RIS } \vspace{0.3cm}\\
P_{\text {control board }}+\left(\mathbb{I}_{v}+\mathbb{I}_{h}\right)\left\lceil\frac{N}{N_{\text{g}} \cdot N_{\text{s}}}\right\rceil\cdot P_{\text {drive circuit }}, \text { varactor-diode-based RIS }
\end{array}\right.
\end{equation}
\hrulefill
\vspace{-0.1cm}
\end{figure*}

Based on the above, $P_{\text {total drive circuits }}$ can be expanded as follows:
\begin{equation}
\begin{aligned}
P_{\text {total drive circuits }} & =N_{\text {drive circuits }} \cdot P_{\text {drive circuit }} \\
& =\left(\left\lceil\frac{N_{\text{c}, v}}{N_{\text{g}, v} \cdot N_{\text{s}}}\right\rceil+\left\lceil\frac{N_{\text{c}, h}}{N_{\text{g}, h} \cdot N_{\text{s}}}\right\rceil\right) \cdot P_{\text { drive circuit }}
\end{aligned}
\end{equation}
 As illustrated in (3), the total power consumption of drive circuits equals the product of the number of driving circuits and the power consumption of a single driving circuit. Particularly, ${{N}_{\text{drive circuits}}}$ is affected by multiple factors, which need to be carefully considered. Firstly, ${{N}_{\text{drive circuits}}}$ is affected by the number of adjustable electronic components, which is expressed as ${{N}_{c,x}}(x=v,h)$. Secondly, ${{N}_{\text{drive circuits}}}$ is affected by the control Degree-of-Freedom (DoF) of RIS, which is expressed as ${{N}_{g,x}}(x=v,h)$. Dividing ${{N}_{c,x}}(x=v,h)$ by ${{N}_{g,x}}(x=v,h)$ gives us the number of control signals we need. Thirdly,  ${{N}_{\text{drive circuits}}}$ is affected by the number of control signals generated by each drive circuit, which is expressed as ${{N}_{s}}$. Dividing $\frac{{{N}_{c,x}}(x=v,h)}{{{N}_{g,x}}(x=v,h)}$ by ${{N}_{s}}$ gives us the number of drive circuits.  At last, the round-up operation $\left\lceil \centerdot  \right\rceil $ is necessary for obtaining an integer value. In addition, ${{N}_{\text{drive circuits}}}$ in both polarization modes (horizontal and vertical) needs to be calculated separately since the RIS unit cell design and DoF control in the two polarization directions may be different.

For PIN-diode-based RIS, $N_{\text {c},x}$ ($x=v,h$)  can be expressed by 
\begin{equation}
 {
N_{\text {c},x}=\sum_{i=1}^N B_{i,x},
}
\end{equation}
where $N$ is the number of RIS unit cells and $B_{i, x}$ is the bit resolution of RIS unit cell in the $x$ polarization direction. Specifically, each PIN diode requires an input control signal to switch between 1 bit-binary state. Thus, $B_{i,x}$ ($x=v, h$) control signals are required for the ${i}$-th unit cell with $B_{i,x}$ bits in the $x$ polarization direction. For varactor-diode-based RIS, a single varactor can be coded to different states by applying different magnitudes of reverse bias voltage to it, thus $N_{\text {c},x}$ can be expressed by 
\begin{equation}
 {
N_{\text {c},x}=\sum_{i=1}^N I_{i,x},
}
\end{equation}
where $I_{i,x}$ is used to indicate that if the ${i}$-th unit cell can be configured in the $x$ polarization direction or not.
For RF switch-based RIS, the ${i}$-th RF switch requires $B_{i,x}$ control signals for $B_{i,x}$-bit coding states.
Therefore, $N_{\text {c},x}$ is the same as PIN-diode-based RISs in (4). In conclusion, $P_{\text {static }}$ is summarized as (6) at the top of this page. 

\begin{table}[]
\renewcommand{\arraystretch}{2}
\caption{{Indicator Parameters of Polarization Mode}}
\centering
\begin{tabular}{|l|l|l|l|l|}
\hline
                        &  {$\mathbb{I}_{v}$} & {$\mathbb{I}_{h}$}  \\ \hline
 {Dual polarization}       & {1}  & {1}  \\ \hline
 {Horizontal polarization} & {0}  & {1}   \\ \hline
 {Vertical polarization}   & {1} & {0}  \\ \hline
\end{tabular}
\end{table}

Overall, the power consumption of the FPGA control board $P_{\text {control board}}$ is modeled as a constant value, and the power consumption of the driver circuits $P_{\text {total drive circuits }}$ is related to the polarization mode, the adjustable electronic components type, the number of adjustable electronic components required, the control DoF, and the number and the power consumption of control signals generated by a single driver circuit. It is worth noting that for \textbf{common RISs}, all unit cells of a RIS are usually set to the same bit resolution, and the control DoF in the two polarization directions is consistent. Therefore, the general $P_{\text {static}}$ in (6) can be simplified to the expression (7) at the top of this page, where $\mathbb{I}_{v}$, $\mathbb{I}_{h}$ are denoted as the indicator parameters of vertical polarization and horizontal polarization respectively, and the parameter “1” indicates that RIS can control incident wireless signals in this polarization direction. The indicator parameters of the polarization mode are shown in Table I.

\subsection{The Power Consumption Modeling of RIS unit cells}

For the family of PIN-diode-based RIS, the power consumption of unit cells needs to be considered thoughtfully. Firstly, it is affected by the polarization mode. More PIN diodes need to be embedded on a dual-polarization RIS to achieve polarizations in both directions at the same time, which will affect power consumption. Secondly, it is affected by the bit resolution of RIS unit cells. Two working states of a single PIN diode can be represented by a 1-bit binary state, while a multi-bit unit cell needs to be realized by multiple PIN diodes, which will affect the power consumption. Thirdly, it is affected by unit cell coding states. For example, a 2-bit RIS unit cell has 4 coding states, which can be denoted as “00”, “01”, “10”, and “11”. The number of “on-state” PIN diodes is different in the 4 coding states, thus the power consumption is also different. 

Specifically, we denote $N$ as the number of RIS unit cells. $P_{i, x}\left(B_{i, x}, b_{i, x}\right)$ ($x=v, h$) is denoted as the power consumption of 
the ${i}$-th unit cell with $B_{i,x}$ bits in the $x$ polarization direction, which can be expressed as
\begin{equation}
P_{i, x}\left(B_{i, x}, b_{i, x}\right)=b_{i, x} \cdot P_{\text{PIN}}, 0 \leq b_{i, x} \leq B_{i, x}
\end{equation}
where $B_{i, x}$ is the bit resolution of RIS unit cell, $b_{i, x}$ is the number of bits encoded as “1”, and $P_{\text{PIN}}$ is the power consumption of PIN diodes for supporting one bit encoded as “1”.
Particularly, if the RIS cannot be configured in the $x$ polarization direction, then $B_{i, x}=0$ and then $b_{i, x}=0$. The expression (8) shows that the power consumption of each unit cell not only depends on the inherent quantization bit $B$ of the unit cell but also on the coding state of the unit cell. For each bit of the unit cell, the power is consumed only when encoded as “1”, and the power consumption is 0 when encoded as “0”.

%====================================================================
\begin{figure*}[!t]
%\large 
\begin{equation}
P_{\text {units }}=\left\{\begin{array}{cl}
\sum_{i=1}^N P_{i, v}\left(B_{i, v}, b_{i, v}\right)+\sum_{i=1}^N P_{i, h}\left(B_{i, h}, b_{i, h}\right), & \text { PIN-diode-based RIS } \vspace{0.3cm}\\
0, & \text { varactor-diode-based RIS } \vspace{0.3cm}\\
\left(\sum_{i=1}^N I_{i, v}+\sum_{i=1}^N I_{i, h}\right) \cdot P_{\text {switch}}, & \text { RF switch-based RIS }
\end{array}\right.
\end{equation}
%\hrulefill
\end{figure*}

\begin{figure*}[!t]
%\large 
\begin{equation}
P_{\text {units }}=\left\{\begin{array}{cl}
\left(\mathbb{I}_{v}+\mathbb{I}_{h}\right) \cdot \sum_{i=1}^N P_i\left(B, b_i\right), & \text { PIN-diode-based RIS } \vspace{0.3cm}\\
0, & \text { varactor-diode-based RIS } \vspace{0.3cm}\\
\left(\mathbb{I}_{v}+\mathbb{I}_{h}\right) \cdot N \cdot P_{\text {switch}}, & \text { RF switch-based RIS }
\end{array}\right.
\end{equation}
\hrulefill
\vspace{-0.1cm}
\end{figure*}
%====================================================================

For the family of varactor-diode-based RISs, the power consumption of unit cells is regarded as 0, since the current in the varactor diodes of the RIS unit cells is negligible when it is working, even though the RIS unit cells are large in number, which is supported by practical measurements\cite{tang2020wireless,9551980}. As for the family of RF switch-based RISs, the power consumption of RIS unit cells is related to the polarization mode, the number of RIS unit cells $N$, and the power consumption of each unit cell $P_{\text {switch}}$. It is irrelevant with the coding state of the RIS unit cells, which is supported by \cite{rossanese2022designing}. Accordingly, $P_{\text {units }}$ can be expressed as (9) on the top of this page. It is a well-known fact that for \textbf{common RISs}, the unit design of all RIS unit cells is the same, which means that all RIS unit cells have the same bit resolution, polarization mode, etc. Therefore, the general $P_{\text {units}}$ in (9) can be simplified to the expression (10) at the top of this page.

Overall, the power consumption of PIN-diode-based RIS unit cells is related to a variety of factors, including the number of RIS unit cells, polarization mode, the precision of regulatory bit, and coding states. The power consumption of varactor-diode-based RIS unit cells is almost zero, and that of RF switch-based RISs has nothing to do with the coding states. Equations (6), and (9) form a \textbf{general RIS power consumption model} together, and equations (7), and (10) form a \textbf{concise RIS power consumption model}.

\section{Practical Measurement and Validation}

In the following section, we assume that all RIS unit cells are set to have the same bit resolution $B$, and the control DoF in the two polarization directions are consistent, expressed as $N_{\text {g}}$. Experimental measurements are carried out to validate the proposed concise power consumption model for RIS, as expressed in (7) and (10). 
Four PIN-diode-based RISs, one varactor-diode-based RIS, and one RF switch-based RIS are utilized for validation, respectively. In the following section, we provide a detailed description of measurement processes and measurement results of the power consumption of practical RISs. 

\subsection{The Practical Measurement of the FPGA Control Board}

\begin{figure}
\centering
\includegraphics[height=2.4cm,width=8.5cm]{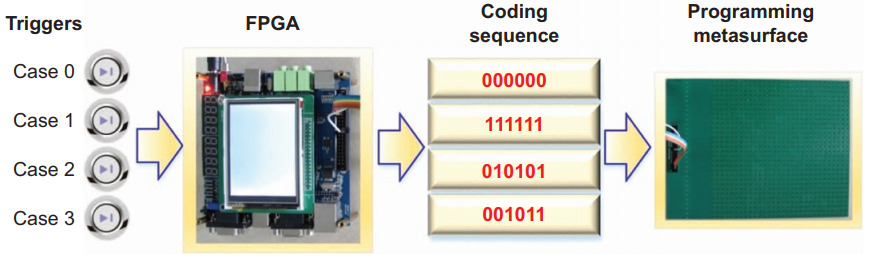}
\caption{A flow diagram for realizing a programmable metasurface controlled by the FPGA hardware.}
\label{r4} 
\end{figure}

Overall, the FPGA control board is like the “brain” of the RIS, which needs to generate the coding sequences for regulating the RIS. A flow diagram for realizing an RIS triggered by the FPGA hardware is presented in Fig. \ref{r4} in \cite{cui2014coding}. In the measured $1^{\#}$ PIN-diode-based RIS and $6^{\#}$ RF switch-based RIS, the XC7K70T is embedded as the master FPGA control board. With a working voltage of 24 V, the current of it is measured as 0.2 A. Therefore, $P_{\text { control board}}=4.8$ W. Some other FPGAs, i.e., Xilinx ZYNQ7100 are also sufficient for RIS prototypes with fewer RIS unit cells or low data processing speed requirements, consuming only 1.5 W like the RIS prototype in \cite{9551980}.

%===========================================
\begin{figure}[!t]
\centering
\includegraphics[height=8.5cm,width=8.5cm]{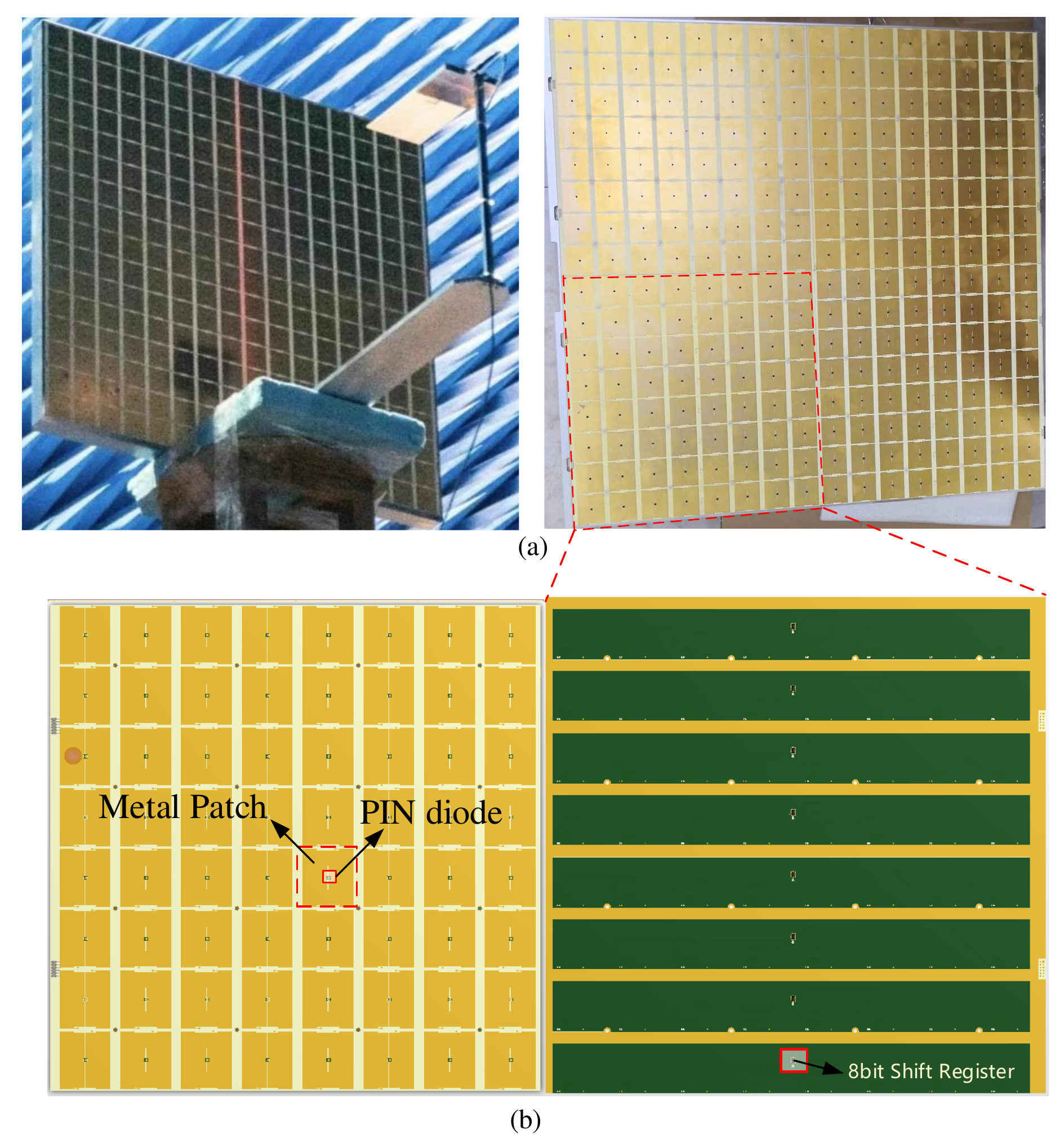}
\caption{Photograph of the fabricated $1^{\#}$ PIN-diode-based RIS. (a) A complete $16 \times 16$ RIS. (b) An $8 \times 8$ sub-RIS structure.}
\label{2} 
\vspace{-0.2cm}
\end{figure}
%===========================================

\subsection{PIN-diode-based RIS}

\subsubsection{Practical Measurement of Drive circuits}
\begin{figure}
\centering
\includegraphics[height=3.3cm,width=8.3cm]{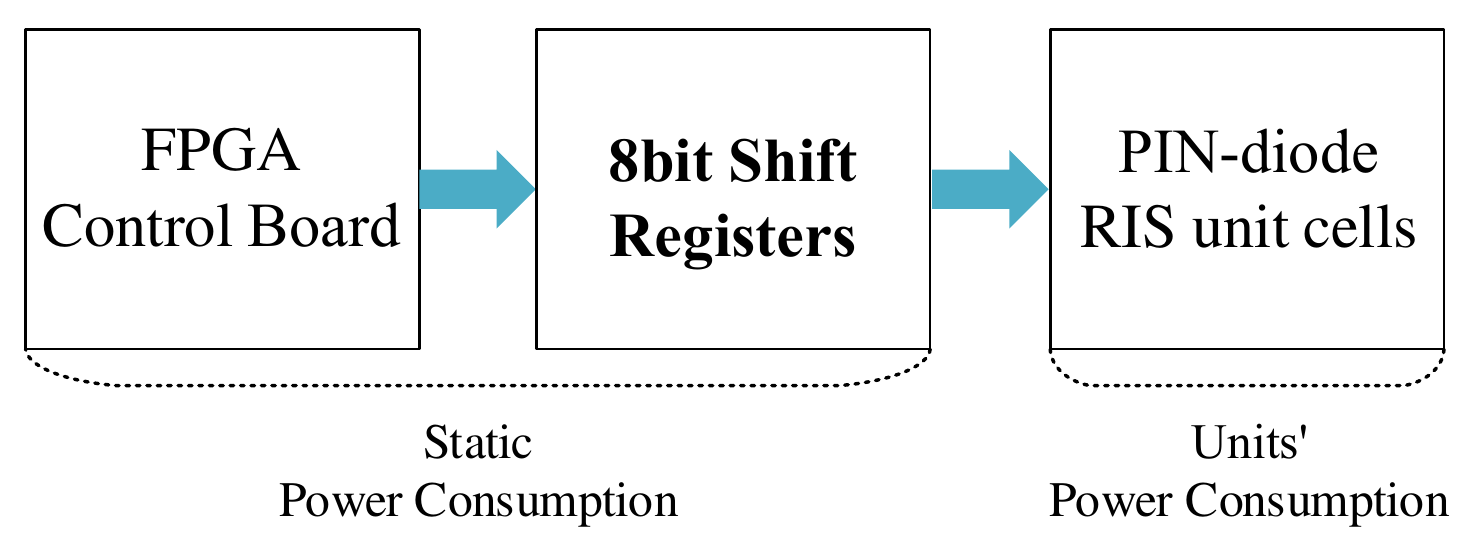}
\caption{The fabricated $1^{\#}$ PIN-diode-based RIS hardware design structure.}
\label{3}
\end{figure}

The $1^{\#}$ RIS utilized for the first measurement belongs to the family of PIN-diode-based RIS, which is phase-programmable with 1-bit coding, as shown in Fig. \ref{2}. It can control the phase shift of reflected EM waves at the operating frequency $f=3.5$ GHz by unit cell independently. Four $8 \times 8$ sub-RISs form a complete $16 \times 16$ RIS. By applying PIN-diode drive circuits, control signals from the FPGA master control board can be changed into output signals with current-driven capability. In the measured PIN-diode-based RIS, the SN74LV595A 8-bit shift registers are applied as drive circuits, as shown in Fig. \ref{3}. 

Specifically, an 8-bit shift register is embedded for 8 PIN-diode-based RIS unit cells on a column in an $8 \times 8$ sub-RIS, which can turn the serial input into multiple parallel outputs. Therefore, $8 \times 4=32$ shift registers are required in the complete $16 \times 16$ RIS. When 8-bit shift registers are working at the voltage $\text{V}_\text {cc} =$ 3.3 V, the current we measured is $\text{I}_\text {cc}  = 20 $ $\mu$A. Thus, $P_{\text {drive circuit }} = 0.07$ mW. The key parameters of this RIS in the model are $P_{\text {control board }} = 4.8$ W, $B=1$, $N=256$, $N_{\text{g}}=1$, $N_{\text{s}}=8$, therefore $N_{\text {drive circuits}}= 32$, and $P_{\text {total drive circuits }}= 32 \times P_{\text {drive circuit }}= 2.24 $ mW. As illustrated in the measurement results, it can be seen that applying shift registers as drive circuits is an energy-efficient solution, ensuring that $P_{\text {total drive circuits }}$ is about mW-level for hundreds/thousands of unit cells.

%===========================================
%===========================================

\vspace{0.3cm}
\subsubsection{Practical Measurement of RIS unit cells}

\begin{figure}
\centering
\includegraphics[height=5.5cm,width=8.5cm]{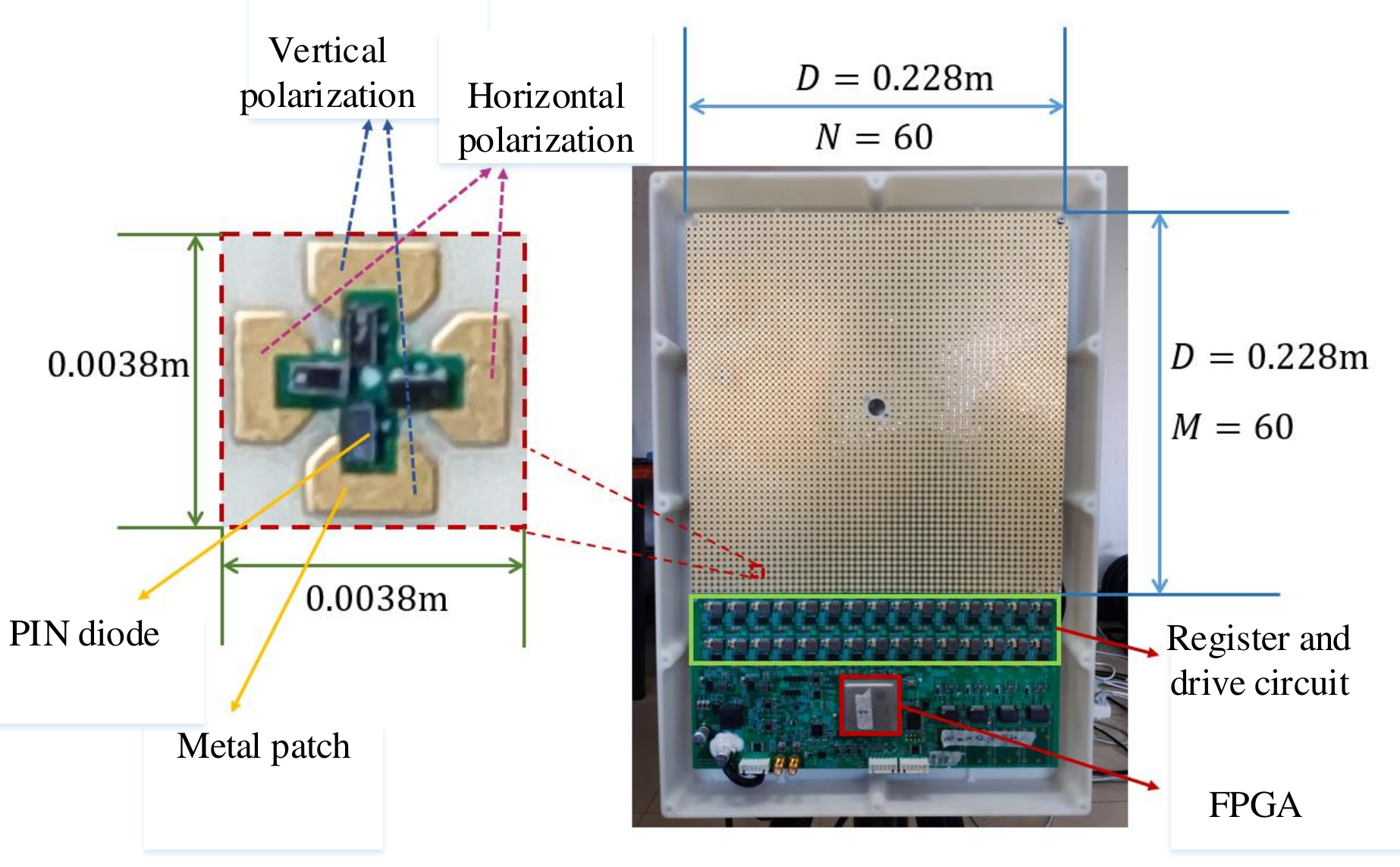}
\caption{Photograph of the $2^{\#}$ dual-polarized PIN-diode based RIS.}
\label{4} 
\vspace{-0.2cm}
\end{figure}

The $2^{\#}$ RIS utilized for verification belongs to the family of PIN-diode-based programmable metasurfaces, which is phase-programmable with 1-bit coding, as shown in Fig. \ref{4}. It is a dual-polarized RIS, which can control the phase shift of reflected EM waves in two polarizations and the operating frequency is $f=35$ GHz. The total length and width of the dual-polarized RIS are $0.228 \text{~m} \times 0.228 \text{~m}$, and the RIS is composed of $60 \times 60$ reduplicated unit cells, with each unit cell having a length and width of $38 \text{~mm} \times 38 \text{~mm}$ and being regulated independently. The unit cell consists of the substrate, two pairs of mutually perpendicular metal patches, and two pairs of PIN diodes connected across the metal patches. For each unit cell, two bias voltages are applied to the two pairs of PIN diodes through metal patches. The bias voltages can change the reﬂection coefﬁcient of the unit cell for vertical polarization and horizontal polarization. 

%==============================================================
%===============================================================

\begin{figure}
\centering
\includegraphics[height=8.7cm,width=8.7cm]{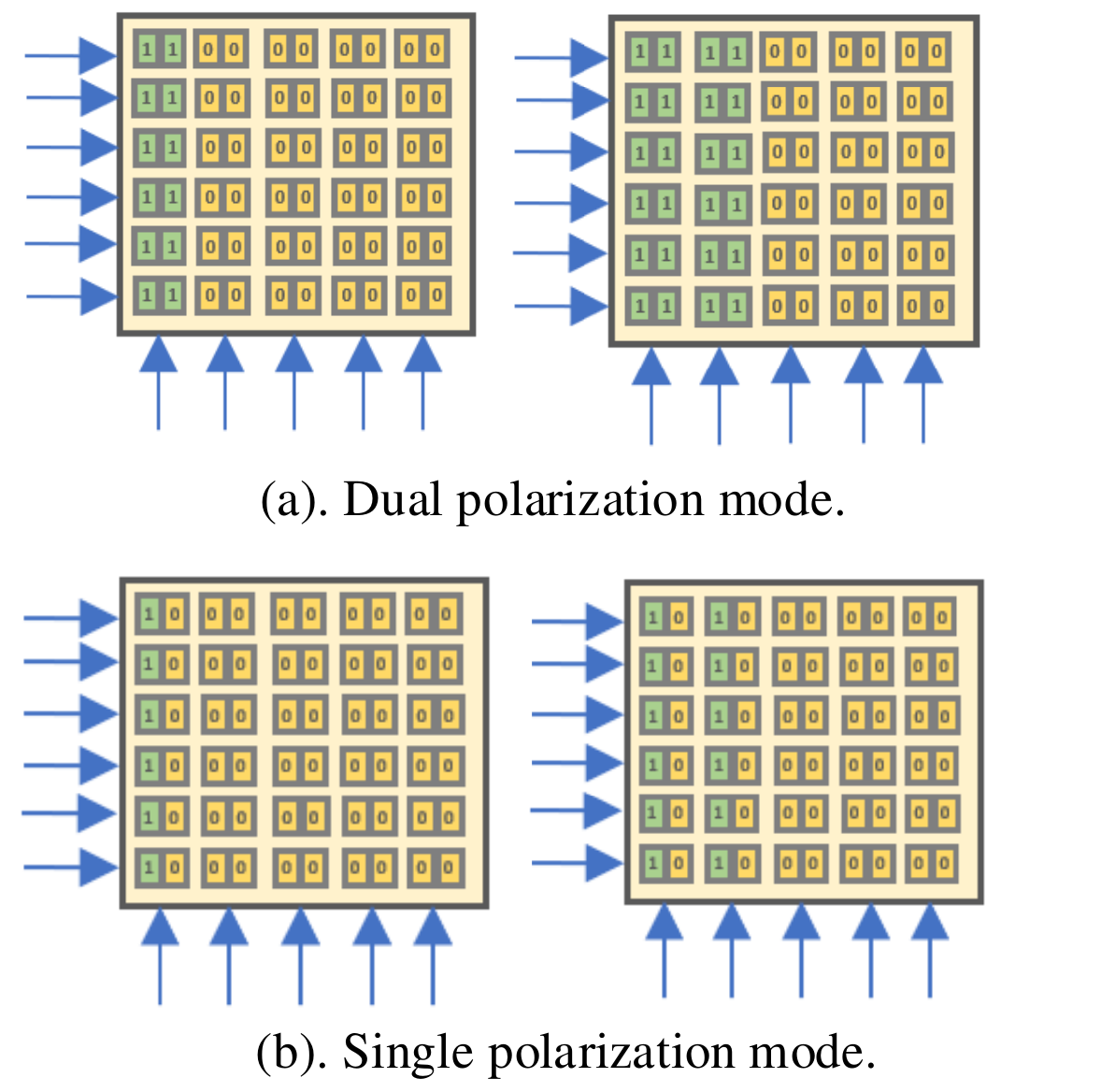}
\caption{Enabling the $2^{\#}$ RIS by column. (a) Dual polarization mode. (b) Single polarization mode.}
\label{5}
\vspace{-0.2cm}
\end{figure}

In the following power consumption measurement, a 0 V voltage is applied to achieve the “0” coding state, and a high (i.e., higher than the conductive voltage of PIN diodes) voltage is applied to achieve the “1” coding state of the RIS unit cell. Firstly, all PIN diodes are encoded as “0”, and all PIN diodes are “off-state”. Therefore, the power consumption at this time is mainly the static power consumption of the control circuit (mainly including FPGA and shift register), measured as 15.75 W.  Secondly, all PIN diodes are encoded as “1”. In this case, all PIN diodes are “on-state” and consume energy, and the maximum power consumption of the PIN-diode-based RIS is measured as 103.2 W. Next, starting from the “0” coding status, each column of the RIS unit cells is switched from “0” to “1”, as shown in Fig. \ref{5}. The power consumption of each measurement is recorded accordingly.

%==============================================================
%===============================================================
%========================================================
\begin{figure}
\centering
\includegraphics[height=6cm,width=9.5cm]{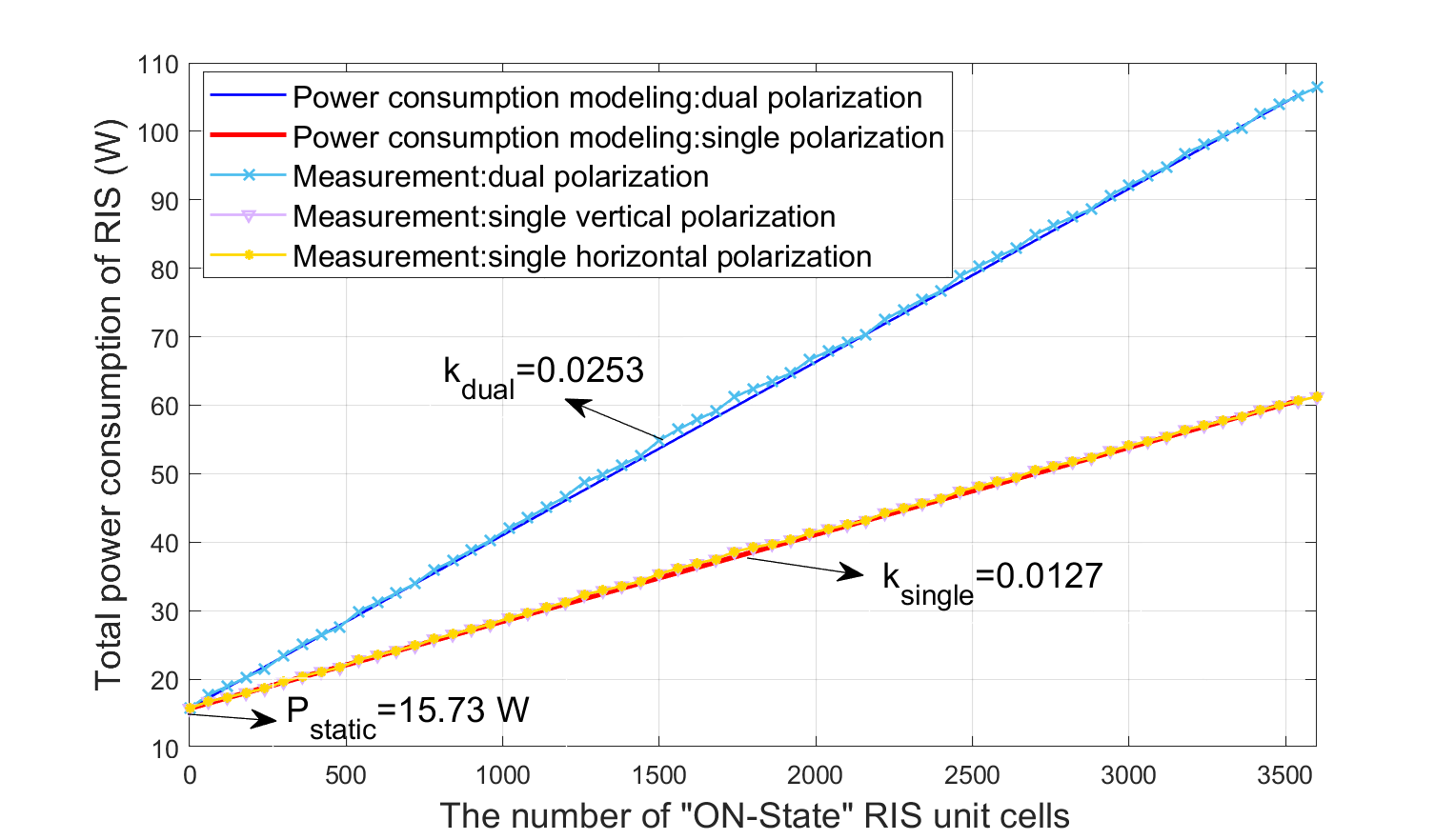}
\caption{The total power consumption vs. The number of RIS unit cells encoded as “1”.}
\label{6}
\vspace{-0.2cm}
\end{figure}
%========================================================
We mainly apply the data fitting method to validate the accuracy and reliability of the power consumption model. For example, when validating the accuracy of the PIN-diode RIS, the RIS utilized for verification is composed of $60\times 60$ unit cells. In single-polarization mode, each column of RIS unit cells is switched from 0 (off-state) to 1 (on-state). In this way, 60 groups of data points (the number of RIS unit cells, power consumption) are recorded. Upon observation, we found that the data distribution conforms to the linear characteristics and, therefore, can be fitted with a linear function $\hat{y}=kx+b$. The slope can be calculated by linear fitting through \textit{Matlab}.

Fig~\ref{6} is to validate the relationship between the total power consumption or RIS and the number of “ON-State” RIS unit cells. When $n=0$, $P_{\text {static }}=15.73$ W, which is the static power consumption; With the increment of $n$, the total power consumption of RIS increases. When $n$ is the same, the power consumption of single vertical/horizontal polarization is approximately equal, and the curves are nearly coincident. Moreover, it can be seen that the power consumption of dual polarization mode is twice as much as that in single polarization mode, denoted as $k_{\text {dual }} \approx 2 k_{\text {single }}$. Therefore, the practical power consumption measurement of $1^{\#}$ RIS is consistent with the proposed model in (10). According to the measurement results, the key parameters of $1^{\#}$ RIS in the model are: $P_{\text {static }}=15.73$ W, $P_{i}(1,0)=0$, and $P_{i}(1,1)=P_{\text{PIN}}=12.56$ mW, $1 \leq i \leq 3600$. As can be seen, the practical measurement results in both polarization modes are in good match with the proposed power consumption model.

\begin{figure}
\centering
\includegraphics[height=8cm,width=9cm]{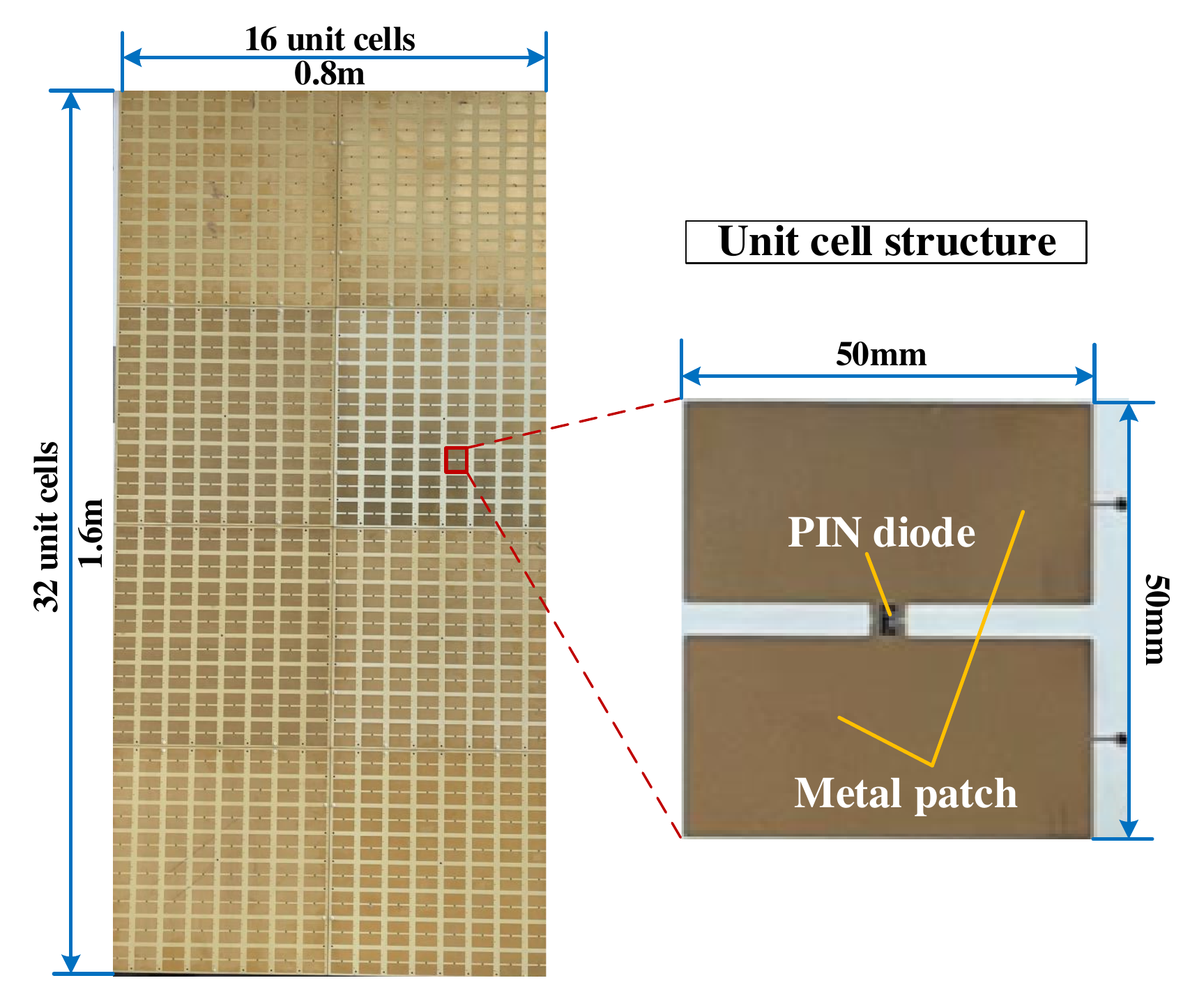}
\caption{Photographs of the $3^\#$ fabricated PIN-diode-based RIS wall.}
\label{7}
\vspace{-0.1cm}
\end{figure}

Another $3^{\#}$ PIN-diode-based RIS utilized for measurement is also a 1-bit coding phase-programmable metasurface, operating at $f=2.6$ GHz, as shown in Fig. \ref{7}. The total length and width of this RIS are $1.6 \text{~m} \times 0.8 \text{~m}$, and the RIS is composed of $32 \times 16$ reduplicated unit cells, with each unit cell having a length and width of $50 \text{~mm} \times 50 \text{~mm}$ and being regulated independently. The overall power consumption of the RIS is 12.66 W when all unit cells are encoded as “1” and 6.52 W when all unit cells are encoded as “0”. 

\begin{figure}
\centering
\includegraphics[height=7cm,width=8.5cm]{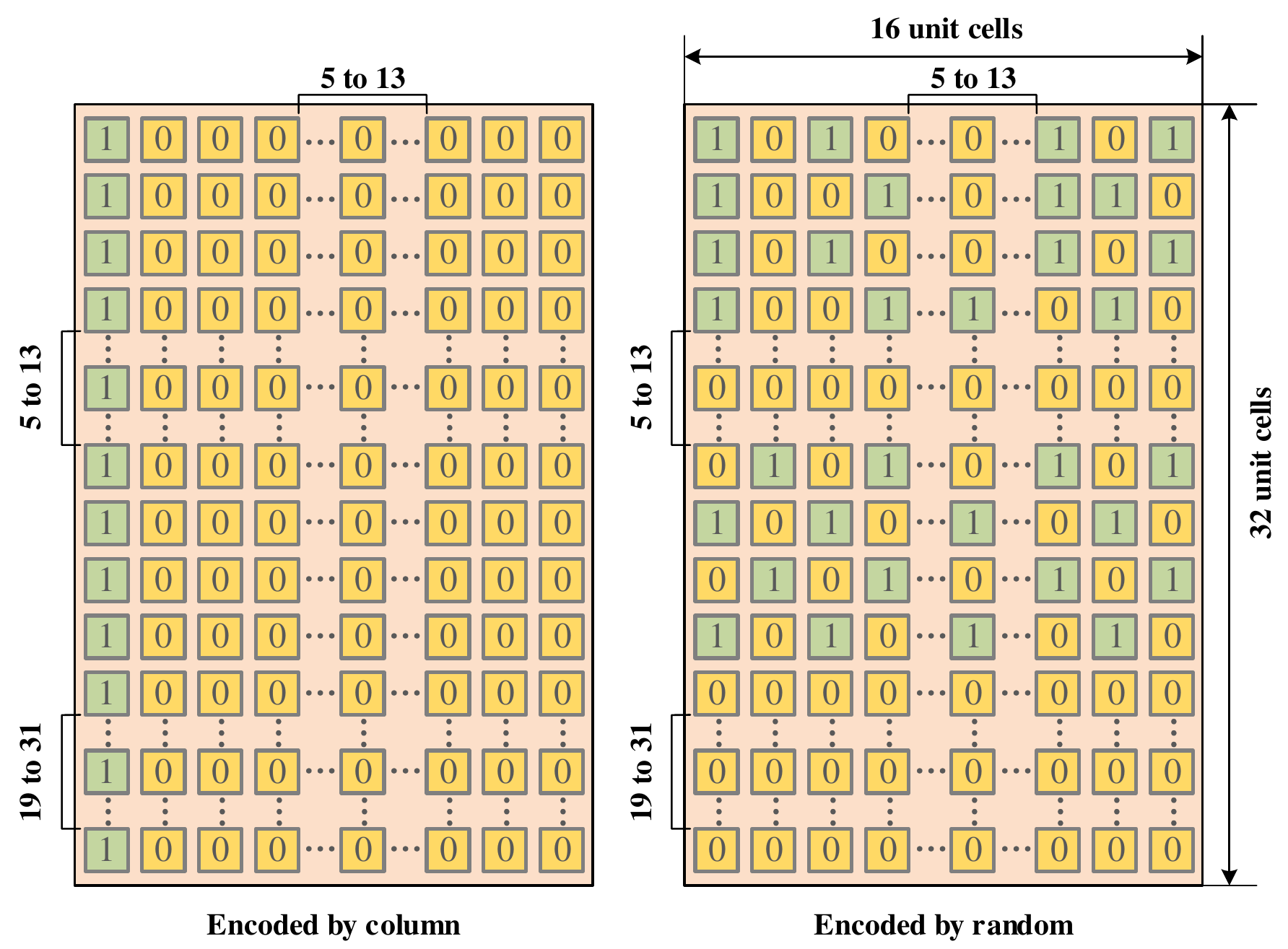}
\caption{Enabling the RIS by column and random.}
\label{8} 
\vspace{-0.1cm}
\end{figure}

This RIS is also encoded from “0” to “1” by column control from left to right and the power consumption is measured and recorded accordingly. The linear relationship between the total power consumption of RIS and the number of RIS unit cells encoded as “1” (denoted as $n$) resembles the first fabricated RIS measurement. When $n=0$, $P_{\text {static }}=6.52$ W, which is the static power consumption; With the increment of $n$, the total power consumption of RIS increases. Moreover, a random encoding measurement is carried out for another model validation. Rather than encoding the RIS by column, the RIS is enabled by random. Encoding the random 32 unit cells as “1” consumes the same power compared with encoding the first column of RIS as “1”, as shown in Fig. \ref{8}. Therefore, the practical power consumption measurement of $3^{\#}$ RIS is consistent with the proposed model in (10). According to the measurement results, the key parameters of $3^{\#}$ RIS in the model are: $P_{\text {static }}=6.52$ W, $P_{i}(1,0)=0$, and $P_{i}(1,1)=P_{\text{PIN}}=11.99$ mW, $1 \leq i \leq 512$.

%========================================================
%========================================================

%========================================================
%========================================================

\begin{figure}
\centering
\includegraphics[height=6cm,width=8.5cm]{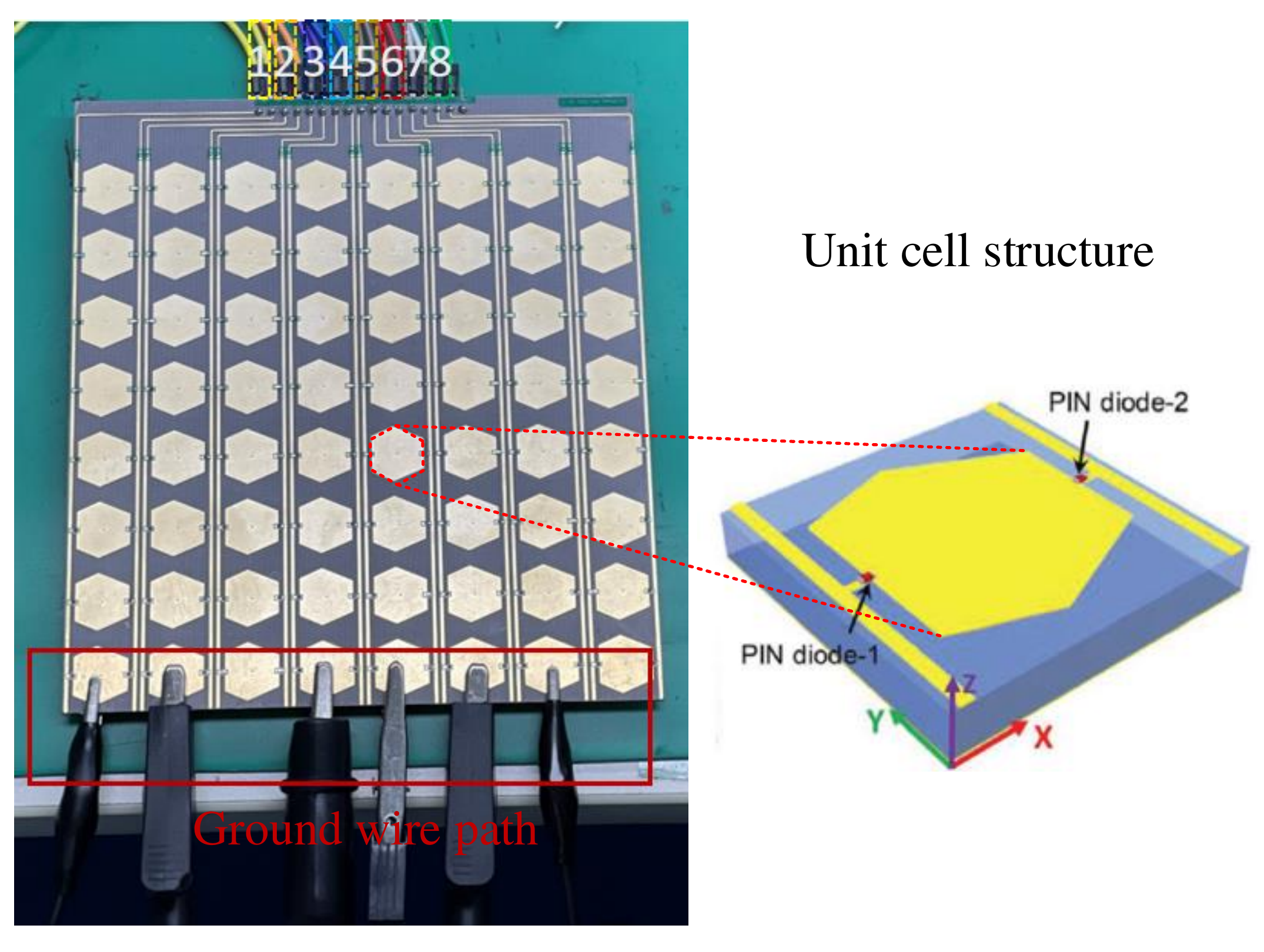}
\caption{Photograph of the $4^\#$ 2-bit RIS.}
\label{9} 
\vspace{-0.1cm}
\end{figure}

The $4^\#$ PIN-diode-based RIS utilized for measurement is a 2-bit coding phase-programmable metasurface \cite{zhang2021wireless,zhang2019breaking}, as shown in Fig. \ref{9}, which is applied to verify the relationship between the power consumption of RIS unit cells and bit resolution of individual RIS unit cell. The 2-bit digital metasurface operates at frequency $f=9.5$ GHz with 8 × 8 programmable elements, and the unit cell consists of a hexagonal metal patch and two biasing lines printed on a grounded dielectric substrate. Two PIN diodes are loaded into each RIS unit cell to control the reflection phase. The size of the RIS unit cell is 14 mm × 14 mm. It is worth mentioning that in the process of this measurement, the ground wire path must be large enough to avoid the nonlinearity caused by the partial voltage of the earth wire resistance. 

\begin{figure}
\centering
\includegraphics[height=6cm,width=9cm]{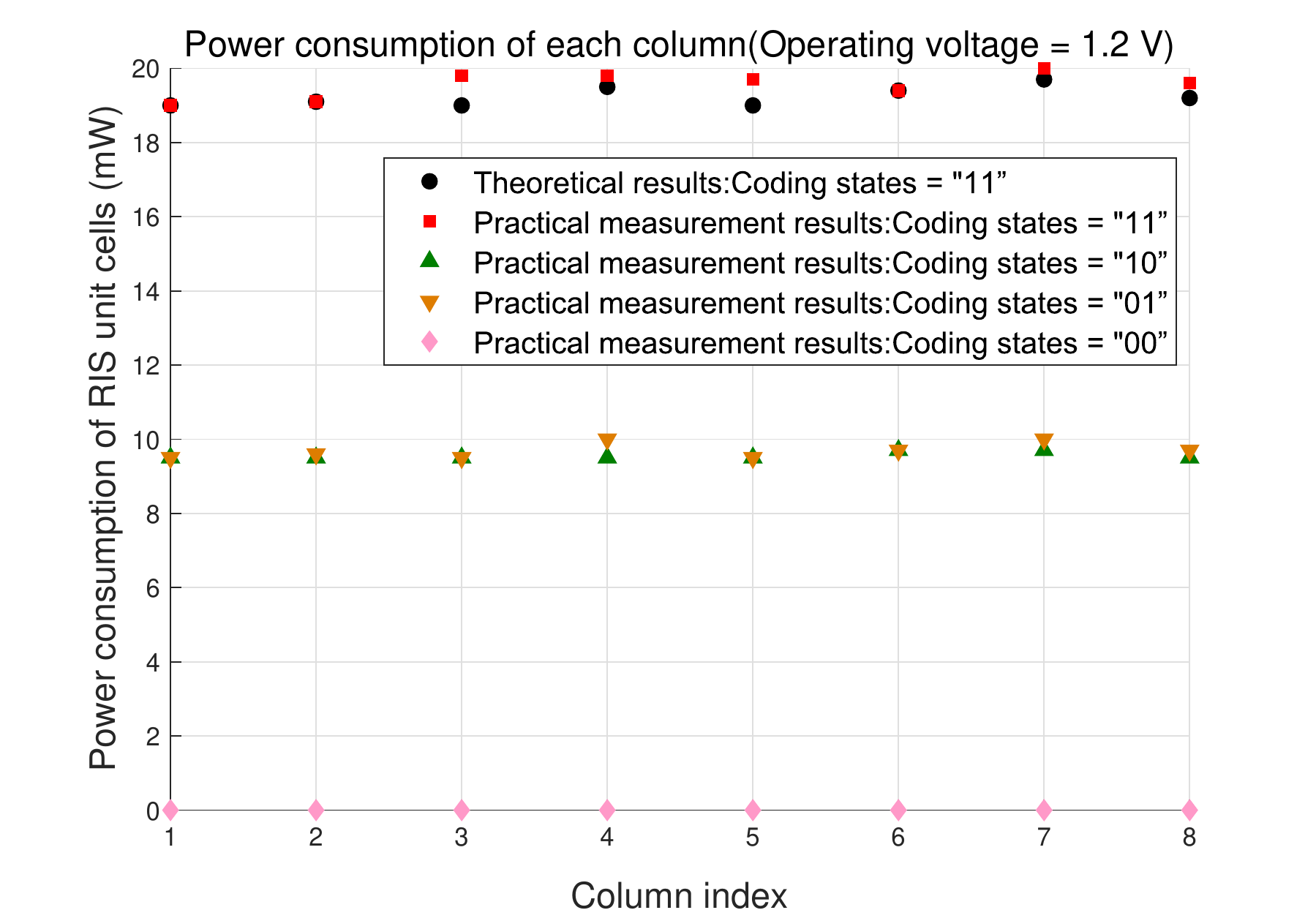}
\caption{Power consumption of each column (Operating voltage = 1.2 V).}
\label{r2} 
\end{figure}

\begin{figure}
\centering
\includegraphics[height=6cm,width=9cm]{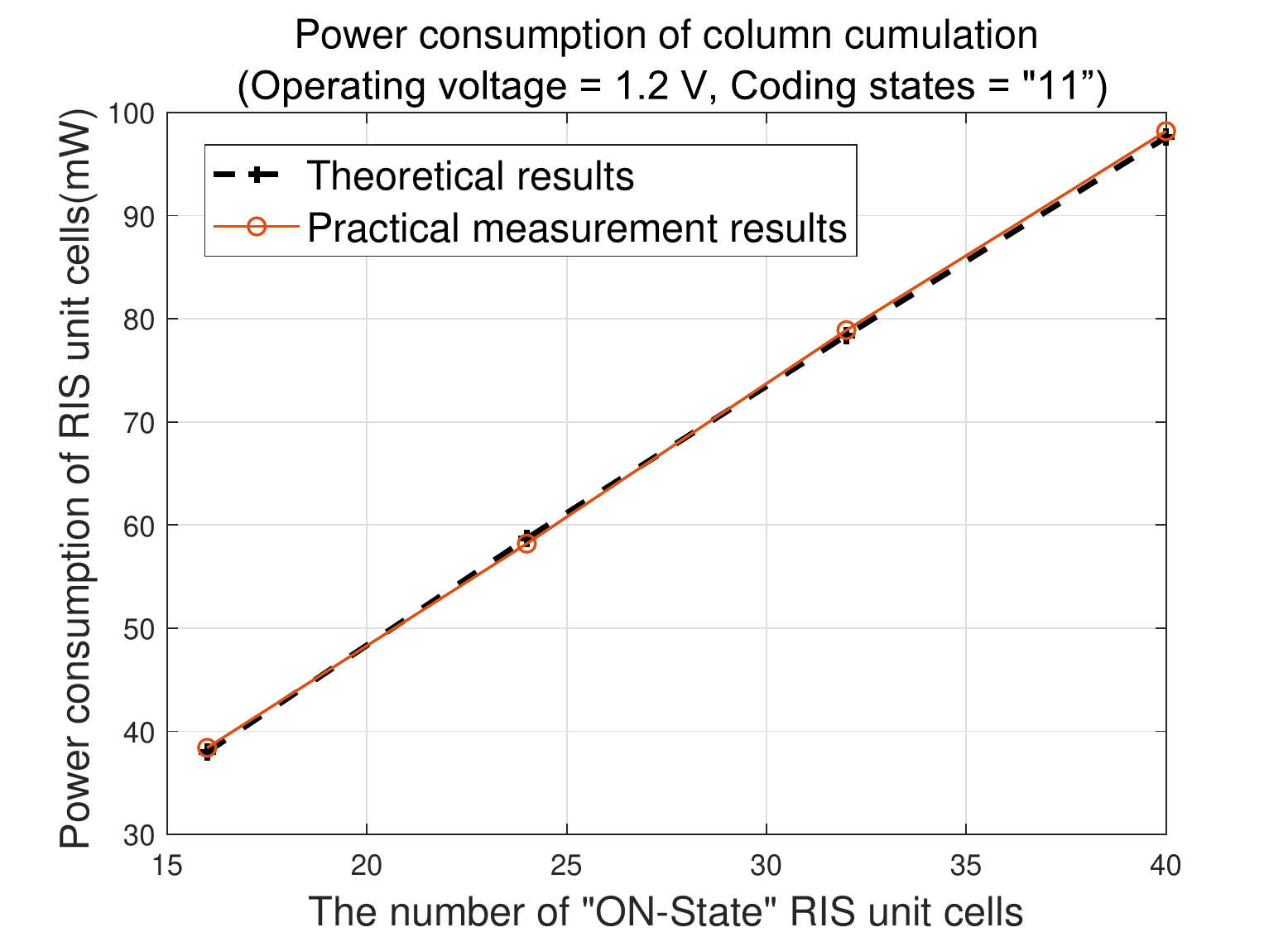}
\caption{Power consumption of column cumulation (Operating voltage=1.2 V, Coding states = “11”).}
\label{r3} 
\end{figure}

When provided 1.2 V voltage, the current of each column is shown in Table II, illustrating that $P_{\text {units}}$ is related to the coding states. Fig~\ref{r2} is to validate the relationship of ${{P}_{\text{units}}}$ and the number of “ON-State” RIS unit cells. The column of RIS is enabled, respectively. And four coding states (“11”, “10”, “01”, and “00”) are set respectively for each column. When provided 1.2 V voltage, the practical measured current of each column with four coding states are shown in the figure.  
Particularly, we add the current of coding states “10” and “01” to get the theoretical current of coding states “11”. As can be seen, the practical measurement results of coding state “11” are in good agreement with the theoretical results of coding state “11”.
When provided 1.2 V voltage with coding states “11”, the power consumption of column cumulation is shown in Fig~\ref{r3} and Table III as well. We enabled 2, 3, 4, and 5 columns (corresponding to 16, 24, 32, and 40 RIS unit cells), respectively. As can be seen, the practical measurement results are still in good match with the theoretical results. According to Tables II and III, we find that there is a linear relationship between $P_{\text {units}}$ and the number of enabled bits of RIS.
Therefore, the practical power consumption measurement of $4^{\#}$ RIS is consistent with the proposed model in (10). According to the measurement results, the key parameters of $4^{\#}$ RIS in the model are: $P_{i}(2,0)=0$, and $P_{i}(2,2)=2P_{i}(2,1)=2P_{\text{PIN}}=2.49$ mW, and $P_{\text{PIN}}=1.25$ mW.

\begin{table*}[!t]
\caption{Power consumption of each column (Operating voltage = 1.2 V)}
\centering
\begin{tabular}{|l|l|l|l|l|l|l|l|l|}
\hline
\makecell[c]{Coding status} & 1 & 2  & 3 & 4 & 5 & 6 & 7 & 8 \\ \hline
\makecell[c]{“11” }   & 19.0 mW & 19.1 mW & 19.8 mW & 19.8 mW & 19.7 mW & 19.4 mW & 20.0 mW & 19.6 mW  \\ \hline
\makecell[c]{“10” }   & 9.5 mW & 9.5 mW & 9.5 mW & 9.5 mW & 9.5 mW & 9.7 mW & 9.7 mW & 9.5 mW  \\ \hline
\makecell[c]{“01” }   & 9.5 mW & 9.6 mW & 9.5 mW & 10.0 mW & 9.5 mW & 9.7 mW & 10.0 mW & 9.7 mW  \\ \hline
\makecell[c]{“00” }   & 0 mW & 0 mW & 0 mW & 0 mW & 0 mW & 0 mW & 0 mW & 0 mW  \\ \hline
\end{tabular}
\vspace{0.3cm}
\end{table*}

\begin{table*}[!t]
\caption{Power consumption of column cumulation (Operating voltage = 1.2 V, Coding states = “11”)}
\centering
\begin{tabular}{|l|l|l|l|l|l|l|l|l|}
\hline
 Columns &1+2 & 1+3  & 1+2+3 & 2+3+4 &1+2+3+4 & 2+3+4+5 & 1+2+4+5+7 & 2+4+5+7+8\\ \hline
 Measured value &38.4 mW & 39.1 mW & 57.5 mW & 58.2 mW &77.4 mW & 77.9 mW & 97.2 mW & 97.6 mW \\ \hline
 Theoretical value &{38.0 mW} & {39.1 mW} & {57.8 mW}& {58.7 mW} &{77.6 mW} & {78.4 mW} & {97.7 mW} & {98.2 mW}\\ \hline

\end{tabular}
\end{table*}

%========================================================

%========================================================

\subsection{Varactor-diode-based RIS}

\subsubsection{Practical Measurement of Drive circuits}

\begin{figure}
\centering
\includegraphics[height=3.1cm,width=8.3cm]{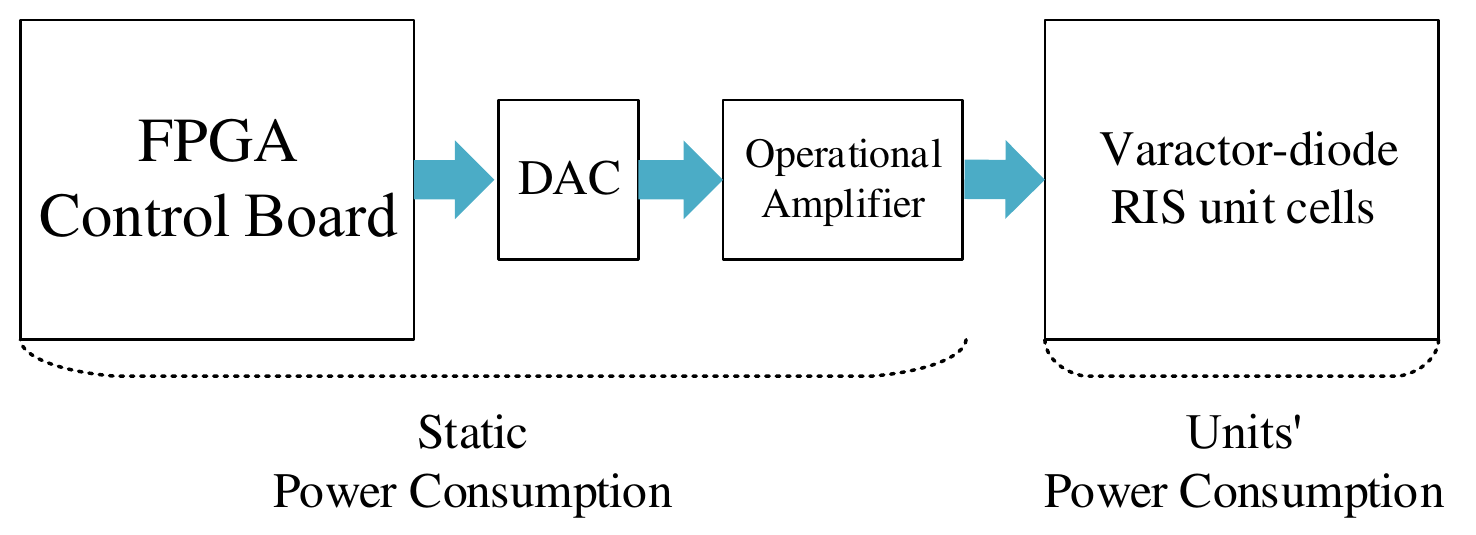}
\caption{The fabricated varactor-diode-based RIS hardware design structure.}
\label{11}
\vspace{-0.2cm}
\end{figure}

The  $5^\#$ RIS utilized for the second measurement belongs to the family of varactor-diode-based RIS, as shown in Fig. \ref{10}(a). It is operated at $f=3.2 $ GHz by column control \cite{liang9632392}. The RIS is composed of $8$ columns with $16$ unit cells in each column. The varactor diode is embedded to bridge the metal patches and operated as an adjustable device and is operated under reverse bias voltages. The typical 8 coding states corresponding to the bias voltages applied on the varactors are illustrated in Fig. \ref{10}(b). Varactor-diode-based RIS unit cells can be driven by DAC + op-amp, PWM signals + level regulator, or CMOS logic circuits. In the measured varactor-diode-based RIS, the DAC3484 and low-noise AD8021 op-amp are applied for continuous phase shift configuration, as shown in Fig. \ref{11}. DACs take digital inputs and generate analog outputs that provide bias voltages of varactor diodes. However, the bias voltage generated by the common DAC is usually low and can not achieve a high one like -20 V in Fig. \ref{10}(b). Therefore, the op-amp is necessary for the operation amplification function. 

In the measurement, 2 columns of RIS unit cells are in the same group with the same control signal, thus $N_{\text {g}}= 16 \times 2 =32$. Specifically, DAC3484 consumes about 250 mW for generating one signal. When the AD8021 works at the voltage $\text{V}_\text {cc}= \pm 12 $ V, the current we measured is 7.5 mA, so the power consumption of the AD8021 is 180 mW. Thus, $P_{\text {drive circuit }}= 250 + 180= 430 $ mW. The key parameters of this RIS in the model are $N_=128$, $N_{\text{g}}=32$, and $N_{\text{s}}=1$, therefore $N_{\text {drive circuits}}= 4$, and $P_{\text {total drive circuits }}= 4 \times P_{\text {drive circuit }} = 1720 $ mW. 

\subsubsection{Practical Measurement of RIS unit cells}

\begin{figure*}
\centering
\includegraphics[height=9cm,width=13.5cm]{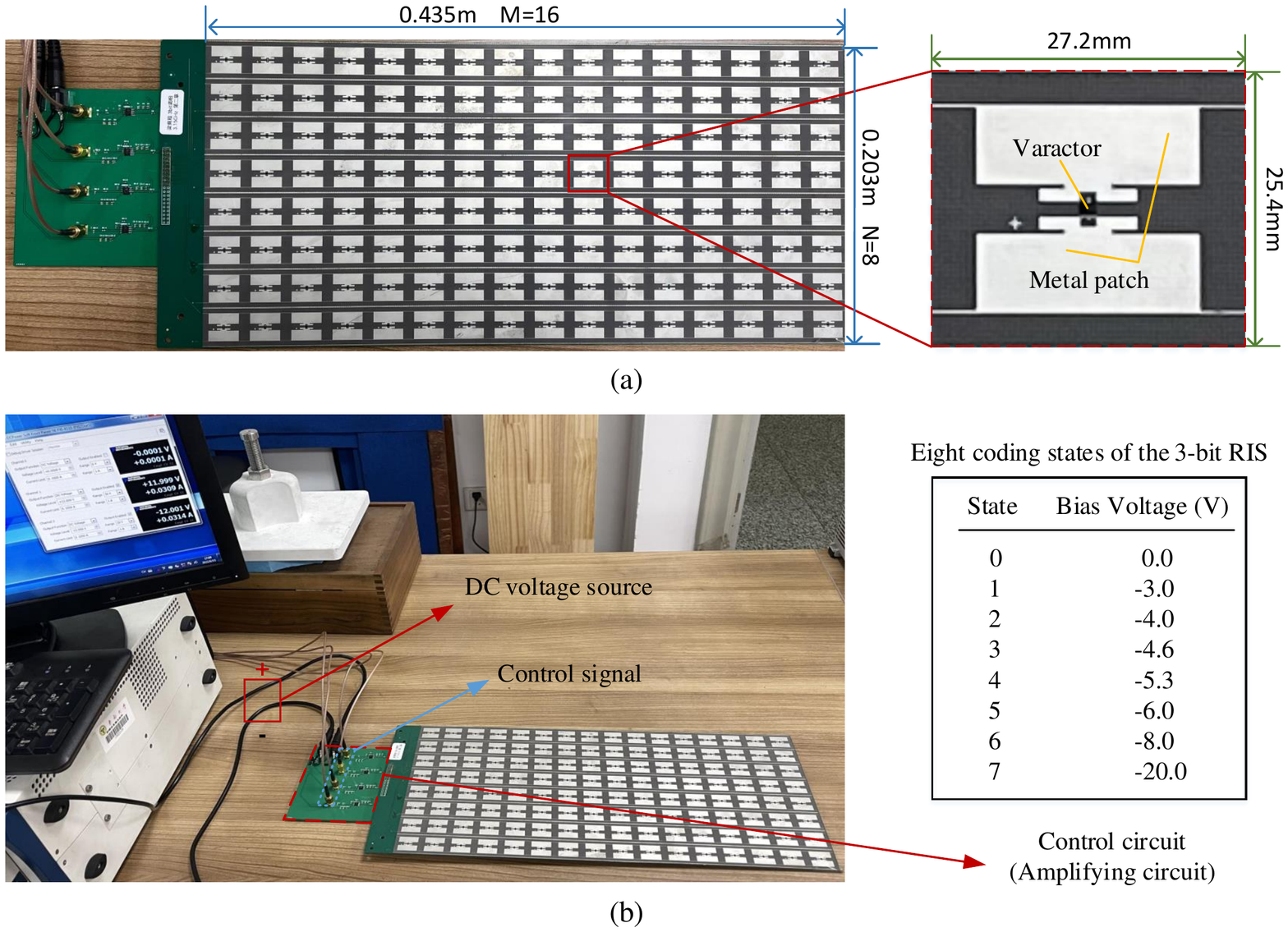}
\caption{Photograph of the fabricated varactor-diode-based RIS. (a) A complete RIS and its unit cell structure. (b) Practical measurement and a typical example of 8 coding states of the RIS.}
\label{10} 
\vspace{-0.2cm}
\end{figure*}

The fabricated $5^\#$ RIS is still utilized for verification. Our measurement setup and 8 coding states corresponding to the bias voltages applied on the varactors are illustrated in Fig. \ref{10}(b). Firstly, 8 reverse bias voltages of $0$ V, $-3$ V, $\cdots$, $-8$ V, and $-20$ V are applied to each column of RIS, respectively. As we expected, according to our proposed model, the current in the varactor diodes of the unit cells is negligible. Next, the RIS is enabled from a single column to all columns, also applying 8 bias voltages, respectively. The current output is also negligible. It is because the junction capacitance changes with the bias voltage only when the reverse bias voltage is applied to the varactor, then the phase modulation function can be realized. The reverse bias voltage will thicken the PN junction inside the varactor and prevent the current from passing through, thus the varactor is always in the “off” state when it works. Therefore, the power consumption of varactor-diode-based RISs is $P_{\text {units}} \approx 0$. 

%==========================================
%===========================================

%===========================================
%===========================================

\subsection{RF Switch-based RIS}

Common solutions implement PIN-diode-based RISs with a finite-bit resolution phase shift and implement varactor-diode-based RISs with both discrete and continuous control of the phase shifts. Recently, RF switch-based RISs have emerged since their implementations entail low costs. Additionally, an RF switch-based RIS may set up individual cells to completely absorb the energy of RF signals in addition to causing phase changes onto impinging signals in a programmed way, which opens up a number of alternatives, just as mentioned in \cite{rossanese2022designing}. 

\begin{figure}
\centering
\includegraphics[height=8.1cm,width=8.1cm]{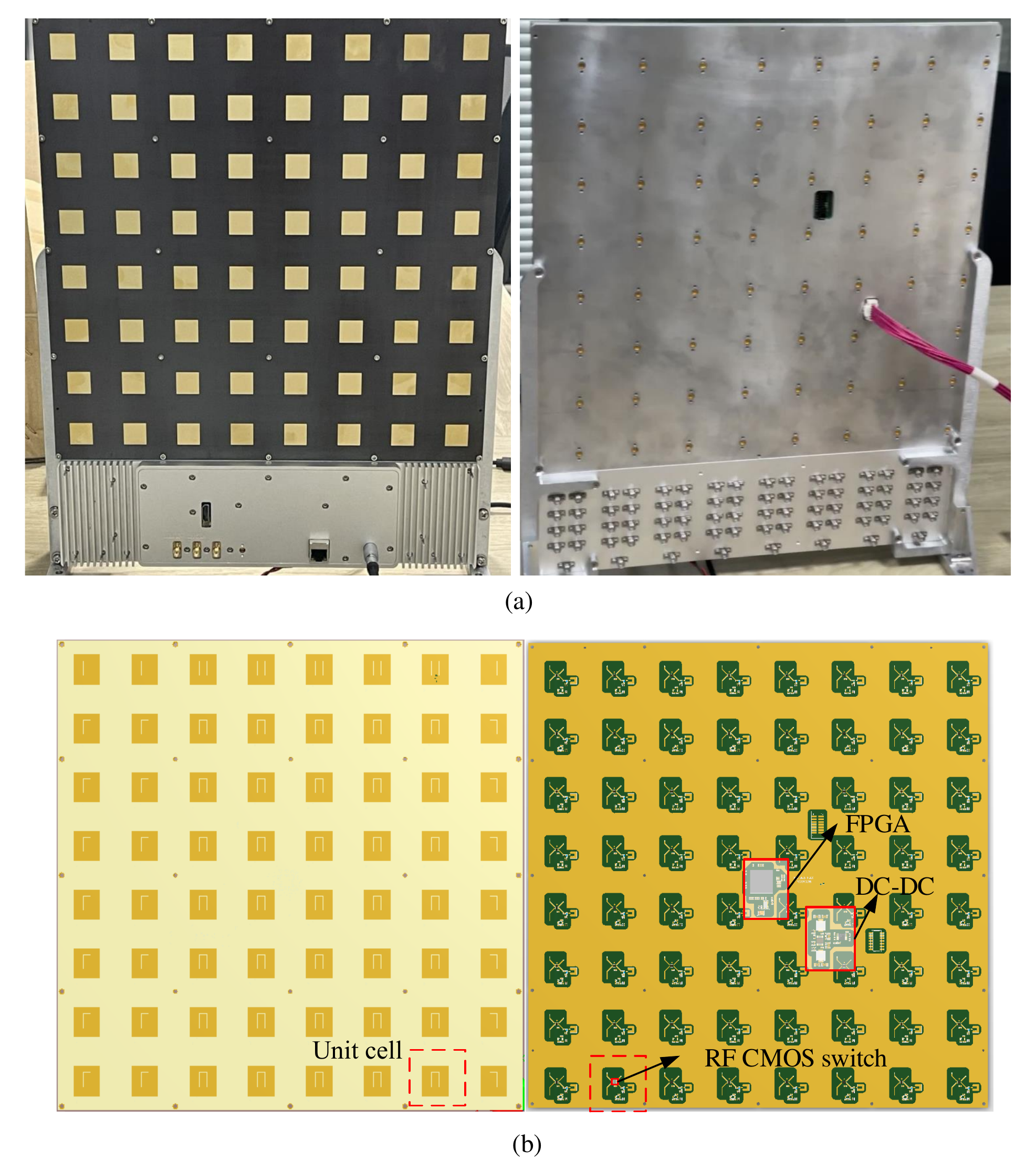}
\caption{Photograph of the $6^\#$ fabricated RF switch-based RIS. (a) The front and back structure of RIS. (b). The physical structure of RIS.}
\label{12}  
\vspace{-0.4cm}
\end{figure}

\begin{figure}
\centering
\includegraphics[height=3.3cm,width=8.3cm]{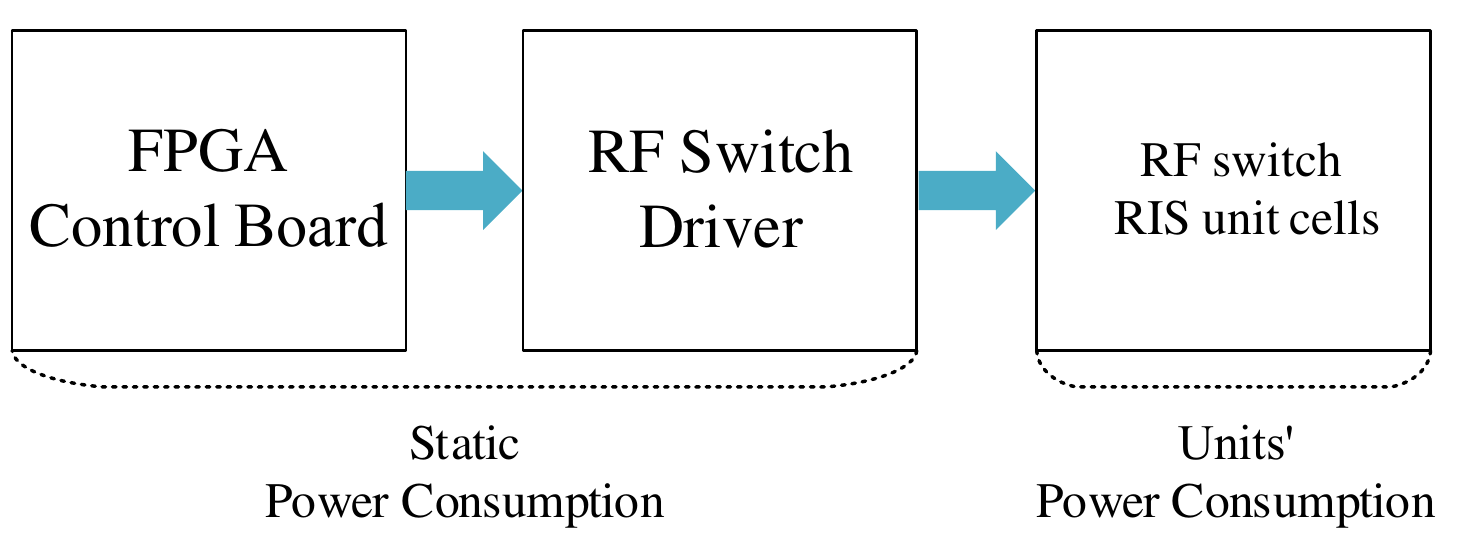}
\caption{The fabricated RF switch-based RIS hardware design.}
\label{13}
\vspace{-0.1cm}
\end{figure}

The $6^\#$ RIS utilized for the last measurement belongs to the family of RF switch-based RIS, as shown in Fig. \ref{12}, and it is composed of $8 \times 8=64$ RIS unit cells, with each unit cell being regulated independently. A RIS unit cell consists of an RF CMOS switch embedded in the metal patch. For an RF switch-based RIS, RF switch drivers are needed to provide logic levels to control the RF switches switching between different coding states, as shown in Fig. \ref{13}. 

In the measured RF switch-based RIS, FPGA XC3S400AN is embedded as an RF switch driver to control 64 RIS unit cells. When XC3S400AN is working at the voltage $\text{V}_\text {cc} =$ 12 V, the current is measured as 20 mA. Therefore, $P_{\text {drive circuit}} = 240 $ mW.  The key parameters of this RIS in the model are $N=64$, $N_{\text{g}}=1$, and $N_{\text{s}}=75$ (a single XC3S400AN can control about 75 RF switches), therefore $N_{\text {drive circuit}}= 1$, and $P_{\text {total drive circuits }}= P_{\text {drive circuit }} = 240$ mW.

\section{Discussion}
\subsection{Discussion about Measurement Results of three types of RISs}
The power consumption of various RISs is now summarized, and the results are all supported by the practical measurements and the product data sheets. First of all, the FPGA control board consumes Watt-level power for data processing. Secondly, even though the $P_{\text {units}}$ of varactor-diode-based RIS is almost zero, the drive circuits of it are often much more complicated and energy-hungry, like $P_{\text {drive circuit }}= 430 $ mW for each control signal with continuous bias voltage variation. It is widely acknowledged that discrete-type varactor-diode-based RISs are promising for energy-efficient design since its $P_{\text {units}}$ is almost zero and its drive circuit $P_{\text { total drive circuits }}$ may sharply decline for only generating discrete-level bias voltage. Despite all this, there still remain challenges for designing and manufacturing the varactor-diode-based RIS in high-frequency bands (e.g., mmWave and Terahertz band). 

In contrast, the drive circuits of PIN-diode-based RIS are energy-efficient, which is mWatt-level consuming, like $P_{\text {drive circuit }}= 0.066 $ mW for each 8-bit shift register and $ 0.066/8= 8 \times 10^{-3} $ mW for each control signal. Nevertheless, the $P_{\text {units}}$ of PIN-diode-based RIS can be very large. The $P_{\text {PIN}}$ of each unit cell in $2^{\#}$ RIS is measured as 12.6 mW, and the maximum $P_{\text {units}}$ is about $90$ W when all unit cells are encoded as “1”. It is noteworthy that $P_{\text{PIN}}$ for $2^{\#}$ and $3^{\#}$ RISs are measured as 12.56 mW and 11.99 mW, respectively, and $P_{\text{PIN}}$ for $4^{\#}$ RIS is measured as 1.25 mW. 
The large power consumption difference between the first two PIN diodes and the last PIN diode is that not only the usage of different PIN-diode components but also the difference in on-voltage applied to the PIN diodes will both lead to the difference in power consumption. When we measured the $4^{\#}$ RIS, the PIN diodes in $4^{\#}$ RIS
the current passing through its PIN diode just reaches its on-state current, thus leading to the low power consumption of the PIN diode. Yet when we measured $2^{\#}$ and $3^{\#}$ RISs, the current on $2^{\#}$ and $3^{\#}$ RIS PIN diodes are greater than the on-state current, which means that sufficient margin is reserved to ensure RISs working properly. 

Furthermore, the $P_{\text {total drive circuits}}$ of RF switch-based RISs is measured as 240 mW for a single drive circuit (FPGA) to control 64 unit cells. Actually, it is entirely possible that the drive circuits for PIN-diode-based RISs and RF switch-based RISs are universal. For instance, shift registers can also be utilized as RF switch drivers. Therefore, $P_{\text {total drive circuits}}$ of RF switch-based RISs is comparable to that of PIN-diode-based RISs. Moreover, $P_{\text {total drive circuits}}$ of RF switch-based RISs can be further reduced for the low power consumption design. For example, the master FPGA control board may directly provide control signals for RF switch-based RIS unit cells, in which way the drive circuits can be omitted, thus $P_{\text {total drive circuits}} = 0$. Additionally, the $P_{\text {units}}$ of RF switch-based RISs is usually ${\mu}$Watt-level power-consuming, about $3.3$ V $\times 150 \mu$A $= 495 \mu$W for each unit cell. 

Overall, PIN diode-based RISs are suitable for high-frequency band (i.e., low-frequency megahertz to mmWave band) scenarios and those scenarios that require discrete phase resolution. Varactor diode-based RISs are suitable for those scenarios that require high phase resolution or precision regulation since varactor diodes can achieve continuous phase shift design. Moreover, the RF switch-based RIS is exactly the most power-efficient among the three types of RISs, keeping low power consumption in both drive circuits and unit cells. Replacing PIN/varactor diode control with new devices or materials, such as RF switch-based RIS, is expected to maintain the RIS phase state with low power consumption. Moreover, the power consumption characteristics of various RISs are summarized in Table IV. 

%==========================================
%===========================================

%===========================================
%===========================================

 \subsection{Discussion about how is the total power consumption modified by the number of RIS unit cells}

\begin{figure*}
\centering
\includegraphics[height=3.7cm,width=18.2cm]{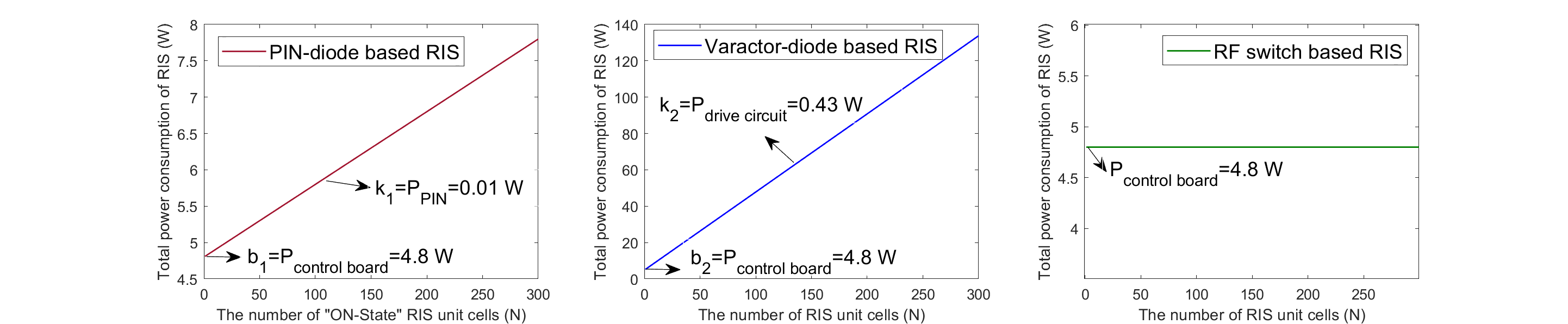}
\caption{The total power consumption vs. the number of RIS unit cells for all three types of RISs.}
\label{r5} 
\vspace{-0.2cm}
\end{figure*}

It is important to discuss how is power consumption modified by the number of RIS unit cells. For PIN-diode based RIS, assuming that it is 1 bit and single polarization, and each unit cell is regulated independently. ${{P}_{\text{control board}}}=4.8\text{ W}$, ${{P}_{\text{PIN}}}=10\text{ mW}$, and ${{P}_{\text{PIN}}}=10\text{ mW}$. ${{P}_{\text{total drive circuits}}}$ can be neglect compared with ${{P}_{\text{units}}}$. Thus, when all RIS unit cells are “ON”, the power consumption of PIN diode-based RIS is simplified to: 
\begin{equation}
{{{P}_{\text{RIS}}}\approx {{P}_{\text{control board}}}+N\times {{P}_{\text{PIN}}} =4.8\text{ +}N\times 0.01\text{ W}}
\end{equation}
For varactor-diode based RIS, assuming that it is single polarization and continuous phase regulation, and each unit cell is regulated independently. ${{P}_{\text{control board}}}=4.8\text{ W}$, ${{P}_{\text{drive circuit}}}=430\text{ mW}$, ${{N}_{g}}=1$, ${{N}_{s}}=1$, and ${{P}_{\text{units}}}=0$. Thus, the power consumption of varactor diode-based RIS is simplified to:
\begin{equation}
{{{P}_{\text{RIS}}}\approx {{P}_{\text{control board}}}+N\times {{P}_{\text{drive circuit}}} =4.8\text{ +}N\times 0.43\text{ W}}
\end{equation}
{For RF switch-based RIS, assuming that it is single polarization and each unit cell is regulated independently. ${{P}_{\text{control board}}}=4.8\text{ W}$, ${{P}_{\text{total drive circuits}}}$ and ${{P}_{\text{units}}}$ can be neglect compared with ${{P}_{\text{control board}}}$. Thus, the power consumption of RF switch-based RIS is simplified to: }
\begin{equation}
{{{P}_{\text{RIS}}}\approx {{P}_{\text{control board}}}=4.8\text{ W}}
\end{equation}
According to the formula (11),(12), and (13), Fig~\ref{r5} shows the power consumption vs. the number of RIS unit cells for all three types of RISs.The figure shows that the power consumption of PIN-diode-based RISs increases with the growth of “ON-State” RIS unit cells. The continuous-type varactor-diode-based RISs are extremely energy-hungry. The power consumption of them increases rapidly with the number of RIS unit cells increasing since the drive circuit of each unit consumes a lot of energy. As for RF switch-based RISs, the number of unit cells has little effect on the power consumption.

 \subsection{Discussion about low power consumption design of RIS-assisted system}
RIS exactly plays an active role subject to optimization with the potential of increasing the system performance of RIS-assisted wireless communication networks. The proposed power consumption modeling of RIS in Section II has revealed that different main factors affect different types of RISs for achieving the low-power-consumption design goals. In the following, we take a single polarization RIS with $10\times10$ unit cells as an example to illustrate the impact of RIS configuration on the system performance.

For the PIN-diode-based RIS, its drive circuits are simple and energy-efficient. $P_{\text {total drive circuits}}$ accounts for a small proportion of $P_{\text {RIS}}$. Therefore, the energy efficiency of the PIN-diode-based RIS-assisted wireless communication system is insensitive to $P_{\text {total drive circuits}}$, more precisely, insensitive to the control DoF of RIS unit cells $N_{\text {g}}$ in (7) in Section II. Particularly, $P_{\text {units}}$ is the essential part that we need to focus on. It is important to stress that the unit cell coding states of the PIN diodes have a very large impact on the total power consumption of RIS. Thus, when addressing the beamforming design of the PIN-diode-based RIS-assisted wireless communication systems, it is necessary to consider the low-power-consumption coding design of RIS unit cells to achieve the high energy efficiency of the entire system. More specifically, the $b_{i}$ in  (10) should be taken as small as possible, which means that we need to make fewer PIN diodes encoded as “1” and more PIN diodes encoded as “0”. We provide a case study of this:  The authors in \cite{practicalmodel} presented a case study on improving RIS energy efficiency. The authors adopted our proposed practical power model for a RIS-assisted multi-user multiple-input single-output (MU-MISO) communication system, which accounts for the difference in power dissipation produced by PIN diode ON-OFF states. Based on our model, a more accurate EE optimization problem is formulated and solved. The case study demonstrates that precisely calculating the power consumption ascribed to PIN diodes is a critical prerequisite when optimizing RIS setups to meet energy-saving needs in RIS-assisted systems.

For the varactor-diode-based RIS, $P_{\text {units}}$ is almost zero, and we only need to pay more attention to $P_{\text {static}}$. And because the drive circuits of varactor-diode-based RIS are often complicated and energy-hungry for
generating continuous phase shifts, thus $P_{\text {total drive circuits}}$ takes a large proportion of $P_{\text {RIS}}$. 
Particularly, it is worth noting that the control DoF of varactor-diode-based RIS $N_{\text {g}}$ has a very large impact on the total power consumption of RIS. We also provide a case study of this: As illustrated in formula (7), if the $10\times10$ RIS ($N=100$) is changed from unit cell control to column control, $N_{\text {g}}$ is changed from $1$ to $10$. Assuming that $N_{\text {g}}=1$, then $P_{\text {total drive circuits}}$ will be reduced to 1/10 of the original value. However, the decrease in the control DoF has a negative impact on the spectral efficiency of the system. Therefore, when addressing the beamforming design of the varactor-diode-based RIS-assisted wireless communication systems, it is necessary to consider the control DoF of RIS unit cells to make a tradeoff between spectrum efficiency and the energy efficiency of the entire system.

As for the RF switch-based RIS, it is exactly a promising type of RIS, keeping low power consumption in both drive circuits and unit cells, for which more investigation and inquiry are expected. It is well worth noting that RISs are expected to be energy efficient, so they are designed to consume as little power as possible. However, some special scenarios (such as beam alignment in mmWave band) could be more power-demanding since a large number of PIN diodes are required for achieving multi-bits, thus maintaining high performance in terms of beam focusing (beam alignment precision). Overall, the low-power-consumption design of RISs is an open research area for further investigation. 

\begin{table*}[!t]
\renewcommand{\arraystretch}{2.3}
\caption{Power consumption characteristics of various RISs}
\centering
\begin{tabular}{|l|l|l|l|l|l|l|l|l|}
\hline
\makecell[c]{{Types of RISs} }& \makecell[c]{$P_{\text {control board}}$}& \makecell[c]{$P_{\text { total drive circuits}}$} & \makecell[c]{$P_{\text {units}}$ } \\ \hline
\makecell[l]{PIN diode-based RIS\\ (Only discrete)} & \makecell[c]{Constant}   &  Simple and energy-efficient
 & Large range changes with RIS unit cells coding  varying  \\ \hline
\makecell[l]{Varactor diode-based RIS\\ (Continuous)}& \makecell[c]{Constant} &  Complicated and energy-hungry
 & Almost zero \\ \hline
 \makecell[l]{Varactor diode-based RIS\\ (Discrete)}& \makecell[c]{Constant} &  Less complicated
 & Almost zero \\ \hline
\makecell[l]{RF switch-based RIS \\ (Only discrete)}& \makecell[c]{Constant} &  Simple and energy-efficient & Fixed lower power consumption\\ \hline
\end{tabular}
\vspace{-0.1cm}
\end{table*}

\vspace{-0.1cm}
\section{Conclusion}
In this paper, we have carried out the power consumption modeling and measurement validation for RISs. According to the measurement results, it can be concluded that RISs have static power consumption $P_{\text {static}}$ and the power consumption of unit cells $P_{\text {units}}$. Particularly, the power consumption of the FPGA control board is regarded as a constant value, however, that of the drive circuit is a variant that is affected by the number of control signals and its self-power consumption characteristics. Moreover, the power consumption of varactor-diode-based RIS unit cells is almost negligible and the power consumption of PIN-diode-based RIS unit cells is related to the polarization mode, controllable bit resolution, working states of RIS, and their quantitative relationship is given in a concise formula which is verified by practical measurements. The power consumption of RF switch-based RIS unit cells is only related to the number of RIS unit cells. The measurement results for each fabricated RIS and the typical value of $P_{\text {static}}$ and $P_{\text {units}}$ are shown and discussed, respectively. Overall, there is room for more study into the low-power-consumption design of RISs.

\ifCLASSOPTIONcaptionsoff
  \newpage
\fi

\vspace{-0.2cm}
 \bibliographystyle{IEEEtran}
 \bibliography{mybib}

\end{document}